\documentclass[12pt, letterpaper]{article}
\usepackage{selectp}
\pdfoutput=1
\usepackage[utf8]{inputenc}

\title{Algorithmic Transparency With Strategic Users}
\author{ Qiaochu Wang \and Yan Huang \and Stefanus Jasin \and  Param Vir Singh\thanks{Qiaochu Wang, Yan Huang and Param Vir Singh are at Carnegie Mellon University. Stefanus Jasin is at University of Michigan.}} 
\date{\vspace{-5ex}}
\usepackage{natbib}
\setcitestyle{comma,authoryear,open={(},close={)}}
\usepackage{graphicx}
\usepackage{xcolor}
\usepackage{bbm}
\usepackage{amsmath,amssymb}
\usepackage{caption}
\usepackage{subcaption}
\usepackage{comment}
\usepackage{stackengine}
\usepackage{soul}
\usepackage[margin=1in]{geometry}

\newcommand\barbelow[1]{\stackunder[1.2pt]{$#1$}{\rule{.8ex}{.075ex}}}
\newtheorem{assumption}{Assumption}
\newtheorem{lemma}{Lemma}
\newtheorem{theorem}{Theorem}

\makeatletter
\newcommand*{\rom}[1]{\expandafter\@slowromancap\romannumeral #1@}

\begin{document}

\begin{titlepage}
\maketitle
\thispagestyle{empty}
\pagestyle{empty}
\begin{abstract}
\noindent Should firms that apply machine learning algorithms in their decision--making make their algorithms \textit{transparent} to the users they affect? Despite growing calls for algorithmic transparency, most firms have kept their algorithms \textit{opaque}, citing potential gaming by users that may negatively affect the algorithm's predictive power. We develop an analytical model to compare firm and user surplus with and without algorithmic transparency in the presence of strategic users and present novel insights. We identify a broad set of conditions under which making the algorithm transparent benefits the firm. We show that, in some cases, even the predictive power of machine learning algorithms may increase if the firm makes them transparent. By contrast, users may not always be better off under algorithmic transparency. The results hold even when the predictive power of the opaque algorithm comes largely from correlational features and the cost for users to improve on them is close to zero. Overall, our results show that firms should not view manipulation by users as bad. Rather, they should use algorithmic transparency as a lever to motivate users to invest in more desirable features.

\vspace{0.1in}

\end{abstract}

\end{titlepage}

\section{Introduction}

Machine learning algorithms are being increasingly applied to decision--making processes with far--reaching impacts extending to employment, access to credit, and education \citep{schellmann2018artificial, wladawsky2019current, fu2018crowd}. However, firms typically keep these algorithms as closely guarded secrets, on par with KFC or Coca-Cola's recipes. As a result, these algorithms stay opaque to the people who they affect and lack clear explanations for the decisions they make. 

Our study is motivated by the growing calls from different parts of society to require firms to make their algorithms transparent. According to American privacy law expert Marc Rotenberg: ``At the intersection of law and technology – knowledge of the algorithm is a fundamental right, a human right.''\footnote{See ``Algorithmic Transparency: End Secret Profiling" https://epic.org/algorithmic-transparency/} The European Union’s General Data Protection Regulation (GDPR) dictates that, whenever personal data is subject to automated decision making, people have ``the right to obtain human intervention on the part of the controller'' or the right to an explanation.\footnote{See ``Algorithmic transparency and the right to explanation: Transparency is only the first step" https://www.apc.org/en/blog/algorithmic-transparency-and-right-explanation-transparency-only-first-step}

While making algorithms transparent is desirable, it can open the door to gaming and potentially adversely affect the classification outcome. If strategic agents were to know the information about the classifier, in other words, how observed attributes affect classification outcome, they may manipulate their attributes to receive a more desirable classification, hurting the predictive power of the algorithm. In financial and economic policy making, this problem is widely known as Goodhart's Law which proclaims that ``when a measure becomes a target, it ceases to be a good measure" \citep{goodhart1984problems}. A similar notion is captured in the Lucas critique \citep{lucas1976econometric}. In fact, Fair Isaac Corporation keeps its exact credit scoring formula secret to make it more difficult for consumers to game the algorithm \citep{Citron}. Similarly, Google continues to modify its secret search ranking algorithm to keep businesses and people from gaming the system \citep{segal2011dirty}.  

Motivated by the calls for algorithmic transparency and the threat of manipulation by the agents to transparent algorithms, we investigate how algorithmic transparency may affect firms and agents. First, from the perspective of the firm (decision--maker) we ask, is there any advantage in making its algorithm transparent when there is the potential for manipulation by agents? Second, we ask, if agents receive more information about the factors affecting algorithmic decisions, would they be better off or worse off if firms make their algorithms transparent? Third, we ask, how are the results affected by the predictive power of those features that are more susceptible to gaming by agents? Finally, we ask, does the market composition in terms of desirable and undesirable agents affect these results? 

We explicitly model agents as strategic and the algorithm designer (the firm) aware of this potential for manipulation. Hence, the firm can react to gaming by the agents. For example, consider that firm collects data, trains an algorithm that maps a set of observed features to a classification outcome, and publishes a decision rule. If agents desire to be positively classified, they would manipulate the values of the features to achieve it. However, the firm would be aware that the behavior of the agents has changed. It will collect new data and update the model and the decision rule. The agents' behavior would change once again. Over time, the firm will iterate to a \emph{fixed point} -- this decision rule would be the best response to the agents' strategic behavior.

More specifically, we model a \emph{job hiring} scenario, where a risk-neutral firm offers a fixed wage and wants to recruit only highly productive agents. There are two types of agents -- High talent (\emph{H}) and Low talent (\emph{L}). The \emph{H} type agents are more productive compared to \emph{L} type agents.  While the type of agent is fixed, this is not observed by the firm ex-ante. However, the firm has access to a number of observed features, \emph{observables}, which the firm uses to map the agent types, using historical data and an algorithm, and determines a decision rule for hiring the agents. 

We model two types of observables, \emph{causal} and \emph{correlational}. Typically in machine learning models, the designer is focused on model accuracy and not on causality. However, any features that are captured by the machine learning model could still be classified as \emph{causal} or \emph{correlational}. For simplicity, we model only two features, one causal and the other correlational. There are several unique characteristics of these features that are important for our model. By definition, the causal feature impacts the productivity of the agent, whereas the correlational feature does not. The agents can game (alter) both features by incurring a cost. As is typically assumed in most signaling game models \citep{Spence1973}, \emph{H} type agents have a cost advantage on the causal feature. We further assume that the cost of improving the correlational feature is type independent and close to zero.

The assumptions behind the cost structures of causal and correlational features warrant some discussion. When an \emph{H} type agent has a significant cost advantage over an \emph{L} type agent regarding the causal feature, it will trivially lead to a separating equilibrium where only \emph{H} types get high values on the causal feature. It is, in cases where the cost advantage of \emph{H} type agents on the causal feature is not large enough, that a firm would want to include the correlational feature in the machine learning model, and hence, in the decision rule. In these cases, the correlational feature would provide additional value in separating the two agent types. It is easy to see that the correlational feature is the one that appears to be the most susceptible to gaming. If agents game it, it will lose its predictive power. This is precisely the reason typically purported for opposing algorithmic transparency. If the cost to alter the correlational feature were very high or the \emph{H} type agent had an advantage over it, either gaming would not happen or gaming would be more favorable to \emph{H} type agents. In such a case, making the algorithm transparent would be either better or at least as good as keeping it opaque for the firm. Our interest is in investigating whether algorithmic transparency can be better for the firm as opposed to keeping the algorithm opaque even when the \emph{H} type agent has no cost advantage on the correlational feature and this cost is close to zero making an algorithm highly susceptible to gaming.

We solve for optimal agent and firm behavior in two scenarios, an \emph{opaque} algorithm scenario and a \emph{transparent} algorithm scenario. In both scenarios, we employ the \emph{Nash equilibrium} as the solution concept. The Nash equilibrium is the fixed--point solution to the game between a strategic firm and agents where the firm (agent)'s action is the best response to the agent(firm)'s action. In equilibrium, neither the firm nor the agents have any incentive to deviate. We then compare the firm's payoff under the two scenarios to determine whether the firm would prefer the transparent or the opaque scenario. 

In the opaque scenario, the agents move first. In this scenario, we assume that the agents are aware of the causal feature but have limited knowledge of the correlational feature. Consequently, the agents can only game the causal feature. Furthermore, the agents know that the firm uses a correlational feature. However, they do not know what that feature is. However, it is common knowledge that the feature has predictive power that can separate \emph{H} type agents from \emph{L} type agents. This assumption is reasonable, as one can see that, if a feature has no predictive power in separating the two types of agents, a machine learning algorithm would discard it.     

In the transparent scenario, the firm moves first and publishes its algorithm. The agents observe this algorithm and know what features are included and their respective weights. In this scenario, the agents can game both the causal and the correlational features. They also know that the correlational feature is correlated with their type.

While in both the opaque and transparent scenarios, the agents move sequentially, we do not employ a \emph{Stackelberg equilibrium} as our solution concept. In the opaque case, given that there are many uncoordinated agents, the unilateral deviation of a single agent will not change the follower (i.e. firm)'s strategy so there might be profitable deviation even if the system reaches a Stackelberg equilibrium. In the transparent scenario, there is a single first mover, and it is possible for the firm to commit to the posted algorithm. Consequently, a Stackelberg equilibrium is a valid solution concept in this scenario. However, a Stackelberg equilibrium provides an advantage to the first mover, which is the firm in the transparent scenario. Thus, the firm could prefer a transparent scenario over an opaque scenario simply due to the first mover advantage it provides. To avoid this possibility, we employ the Nash equilibrium as the solution concept in both the transparent and opaque scenarios.  
`

Our first result challenges the conventional wisdom that making algorithms transparent will always hurt the firm (decision maker) economically. We \textbf{identify a broad set of conditions under which making the algorithm transparent is beneficial for the firm}. The key intuition behind this result is driven by how \emph{H} type and \emph{L} type agents respond to algorithmic transparency. Because investment into the causal feature is costly and because $H$ type agents have advantage on the correlational feature, they invest in improving the causal feature only to the extent that it, along with the correlational feature, helps them separate themselves from \emph{L} type agents. When the algorithm is made transparent, the \emph{L} type agents game the correlational feature. As a result, \emph{L} type agents become similar to \emph{H} type agents on that feature, and the predictive power of the correlational feature decreases. Hence, the \emph{H} type agents have to invest more in the causal feature to separate themselves from the \emph{L} type agents. When the \emph{H} type agents have significant cost advantage over the \emph{L} type agents on the causal feature, this leads to full separation, which benefits the firm. When the cost advantage of the \emph{H} type agents on the causal feature is marginal, the \emph{L} type agents also invest significantly into the causal feature once the disadvantage they faced due to the correlational feature disappears under transparency. In this case, both the \emph{H} and \emph{L} type agents become more productive because of higher investment in the causal feature.  Even though the firm cannot separate the two types of agents in this case, when the causal feature's impact on productivity is above a certain threshold, the average productivity of the hired agents is significantly higher than in the opaque case. In other words, making the algorithm transparent allows the firm to motivate the agents to invest in improving features that are valuable to the firm.

Agents have more information under the transparent scenario. Consequently, one would think agents would be better off under the transparent scenario. However, our second result is that \textbf{the agents are not always better off under the transparent scenario.} There are conditions under which the agents are worse off in the transparent scenario. More interestingly, in most cases where the firm prefers the transparent scenario, the agents prefer the opaque scenario and vice versa. However, we also identified a set of conditions where both the firm and the agents prefer the transparent scenario. The intuition for this result is similar to that of the first result. The firm prefers the transparent scenario in situations where transparency motivates the agents to invest highly in the causal feature. When this cost of investment is high and the \emph{H} type agents have significant advantage, only the \emph{H} type agents will invest in the causal feature. In this situation, although only the \emph{H} type agents are hired, they are worse off because of the high cost that they incurred. When this investment cost is moderate  and the \emph{H} type agents have a marginal cost advantage over the \emph{L} type agents, both agent types invest in the causal feature, and both are hired. The agents are better off because they have to incur a moderate cost for being hired. Simultaneously, the firm is better off as the average productivity of the hired agents is higher when the impact of causal feature on productivity is above a threshold.

As the predictive power of the correlational feature increases, one would expect the firm to be better off with an opaque algorithm. However, our third result shows it is possible that \textbf{the firm prefers algorithmic transparency when the correlational feature has high predictive power and prefers an opaque algorithm when the correlational feature has low predicted power.} The key intuition behind this result is that, when the correlational feature is good at separating \emph{H} type agents from \emph{L} type agents, the \emph{H} type agents have little incentive to invest in the causal feature in the opaque scenario. However, under transparency, the \emph{H} type agents lose this advantage and have to invest in the causal feature to separate themselves from \emph{L} type agents. As a result, the firm can hire more productive agents under transparency.  

Our fourth result is that, \textbf{when the fraction of \emph{H} type agents on the market is higher, the firm prefers the transparent algorithm under certain conditions.} More specifically, the fraction of \emph{H} type agents impacts the firm's surplus but does not have a large enough impact to alter its decision for or against transparency when the cost for improving the causal feature is too high or too low or the \emph{H} type agents have a large cost advantage over the \emph{L} type agents. However, when the cost as well as the cost advantage for the \emph{H} type agents is moderate, the firm prefers algorithmic transparency as the fraction of \emph{H} type agents on the market increases. In this cost range, both the \emph{H} and \emph{L} type agents would improve on the causal feature. While the firm is unable to separate the two and hires both, the number of \emph{L} type agents that are hired becomes smaller as the fraction of \emph{H} type agents on the market increases. 

Our paper makes several contributions. This paper is one of the first to provide an analytical model that systematically compares a firm's decision choice of algorithmic transparency \emph{versus} opacity in the presence of strategic agents. We show that, counter to conventional wisdom, the firm could be better off under algorithmic transparency. By contrast, in most cases where the firm prefers algorithmic transparency, the agents will be worse off. Agents underinvest in the causal feature when the algorithm is opaque. Consequently, the firm depends upon the correlational feature to separate them from one another. Our results and analysis show that the firm should not worry about the potential loss of predictive power of the correlational feature of its machine learning model under transparency. Rather, it should use algorithmic transparency as a lever to motivate the agents to invest more in the causal feature. The firm would typically be reluctant to adopt algorithmic transparency when their machine learning model derives large predictive power from the correlational feature. However, we show that the firm should recognize that investment in the causal feature by agents is endogenous. Agents are less likely to invest in the causal feature when the correlational feature is able to separate them. This is the scenario where the firm should be willing to lose the predictive power of the correlational feature. We have demonstrated our results in conditions where the firm is certain it will lose the predictive power of the correlational feature and does not have a first mover advantage under algorithmic transparency or the difference in the causal feature is insufficient for separating the \emph{H} type and \emph{L} type agents in the opaque machine learning model. Intuitively, one would think that algorithmic transparency would be bad for the firm under these conditions. We have shown that, once one considers endogenous investment in the causal feature, the competition between the \emph{H} and \emph{L} type agents, and the impact of the causal feature on productivity, the firm would be better off making the algorithm transparent.  

The rest of the paper is organized as follows. We discuss how we build upon and contribute to the literature in Section \ref{Literature}. The main model is presented in Section \ref{Model}. In Section \ref{Analysis}, we present the analysis of the model, and we conclude the paper in Section \ref{Conclusion}. 

\vspace{-0.3cm}
\section{Literature}\label{Literature}

Algorithmic transparency is a relatively new topic, but it is closely related to literature on information asymmetry. Following the canonical job market signaling model developed by \cite{Spence1973}, a rich stream of research has focused on the interaction between a decision maker and strategic agents under asymmetric information. Some of this research is focused on the agent's side and study how agents strategically reveal their type under various market conditions (e.g., the stream of signaling game literature). Other research is focused on the decision makers' side, studying how they can design optimal algorithm to extract agents' private information (e.g., the stream of strategic classification literature). Our work on algorithmic transparency is built upon and contributes to both these streams of literature. 

 In both opaque and transparent scenarios, the interaction between the firm and the agents can be adapted into the signaling game framework, where individuals first send signals, and the firm then makes hiring decisions based on the observed signals \citep{Spence1973,engers1987,weiss1983sorting,DALEY2014}. The signaling game models focus on specifying the equilibrium outcome under various market conditions. Receivers are assumed to be in a competitive environment and always receive zero profit in the equilibrium. All the surplus is extracted by senders. The equilibrium concept typically used in signaling game models is the perfect Bayesian equilibrium (PBE), where three conditions are satisfied: senders are using optimal strategies facing the wage offer, receivers give wage offers such that they will obtain zero profit, and the receiver's beliefs about the senders' type given the signal is consistent with the truth. Similar to the signaling games, we also specify equilibrium under various market conditions under the transparent and opaque scenarios. However, in our model, the firm offers a fixed wage and is focused on designing the algorithm to increase its chances of hiring the most productive agents, which contrasts with the signaling games where all agents are hired and the firm's objective is to decide how much salary to offer to each agent,
 
 On the firm side, our model setup bears more similarity to the strategic classification problems -- offering a fixed wage, the firm decides whether to hire the agents based on the signal \citep{kleinberg2018,frankel2019, Bonatti2019}. In other words, the firm is trying to classify agents based on whether their expected productivity exceeds the wage or not. This classification setting makes it possible for us to analyze the economic impact of algorithmic transparency on the decision maker by comparing their equilibrium payoff in the opaque and transparent scenarios. Moreover, we do not assume the signal (education level in our example) to always be pure money burning. Instead, we allow the causal feature to positively impact productivity and we specify conditions regarding this positive effect under which a transparent algorithm will benefit the decision maker. In that aspect, our work is also closely related to the strategic classification literature that assumes the existence of both causal and non-causal features. In contrast to the strategic classification literature that usually employs the Stackelberg equilibrium, we use the Nash equilibrium for both the opaque and transparent scenarios. Several related studies have adopted the Nash equilibrium as the solution concept \citep{bruckner2012, adversarialclassification}. More recently, research in both economics and computer science has noted that under Stackelberg equilibrium, the decision maker can receive a higher payoff compared to Nash equilibrium, but this may result in heavier social burdens \citep{frankel2019, milli2018}.

\textbf{Signaling Games:} Signaling game literature studies how agents strategically reveal their type to a principle in a situation of information asymmetry. Traditional signaling models typically assume that costly actions are the only channels through which agents can signal their type \citep{Spence1973}. In these models, standard assumptions such as the Spence-Mirrlees single-crossing condition ensure the existence of separating equilibria: equilibria that fully reveal agents’ private information. While the machine learning models are trying to solve the same problem (i.e., a decision maker trying to identify the type of agents under information asymmetry), they differ from the classical signaling models in a number of ways. A machine learning model uses multiple features through which it tries to learn an agent's type. Each feature is essentially an action taken by the agent that signal their type. Some of these features are costly, while others are not. 

Our paper is related to recent signaling papers that have also considered multiple actions as channels through which an agent can signal their type \citep{engers1987, frankel2019, DALEY2014, ALOSFERRER2012}. In these papers, the agent is always aware of the actions that are used as signals by the decision maker. In contrast, in our model's opaque case, the agents know that a correlational feature is being used by the firm, but they do not know exactly what feature that is.

In our model, agents can use causal and/or correlational features to signal their type. The causal feature is similar to the costly signal dimension typically captured in the traditional signaling literature. A key difference is that we allow it to act as a signal of an agent's type as well as have an impact on productivity, similar to \cite{weiss1983sorting}. For example, if agents of the same type have different levels of education, in our model, the firm would receive different values from them. The correlational feature that we model bears some similarities to the information in cheap talk games \citep{crawford1982strategic}. This feature is almost costless to share, and it affects the eventual payoff of both the firm and the agent where their incentives are not perfectly aligned. In cheap talk games, the agent strategically manipulates this information, whereas, in our model, the agent does not know about this feature and cannot manipulate it in the opaque case.   

Similar to our model, a few recent papers have modeled the trade-offs that an agent faces in the presence of multiple signals. \cite{DALEY2014} model a scenario where a student can send a costly signal (e.g., joint degree completion) or rely on a type--correlated noisy signal (e.g., grades) to the recruiter. They characterized the results in terms of the informativeness of the noisy signal. The noisy signal is similar to the correlated feature that we model, and its informativeness is also modeled similarly to how we capture the predictive power of the correlated feature. A key finding of this paper is that, when grades are informative, \emph{H} type senders are less eager to send costly signals because they can now rely on grades, while \emph{L} type senders are more willing to send signals to de-emphasize the grades dimension. Consequently, separating equilibrium on the costly signal dimension is harder to sustain. Our model also shares some similarities with this paper in that we also point out the possibility that this extra informative dimension will change individuals' behavior on the more costly signaling dimension. However, there are key differences in our model's assumptions and results. Unlike the grades dimension, where individuals have little chance to manipulate its value, in our model, the firm can give individuals the opportunity to game this correlational feature by making the algorithm transparent. On the one hand this dimension will be less informative. On the other hand, individuals' behavior on the more costly signaling dimension will also change and could lead to a separating equilibrium in many cases. 


\textbf{Strategic Classification:} Strategic classification literature considers the problem of designing optimal classification algorithms when facing strategic users who may manipulate the input to the system at a cost \citep{Hardt}. Canonical strategic classification models deem that the user's manipulation always hurts the decision maker. Guided by this belief, a large stream of research on strategic classification is focused on developing algorithms that are robust to gaming \citep{meir2012,cummings2015,chen2018}. Recently, several papers have argued that this gaming itself can be beneficial to the decision maker; thus, instead of focusing on manipulation-proof algorithms, these papers focus on designing algorithms that incentivize individuals to invest in desirable features \citep{kleinberg2018,Alon2020,haghtalab2020}. These papers are the ones we want to highlight since our paper also points out the difference between `gaming' and  `improvement': gaming is bad for the decision maker because it deteriorates the information contained in the relevant features, but `improvement' could be beneficial to the decision maker since it will causally impact the target variable.

\cite{kleinberg2018} studied the principle-agent problem where the agents' features (e.g., final exam score) can be improved in two ways: by investing effort in a desirable way (e.g., spending time on course material) or by investing effort in an undesirable way (e.g., cheating). The effectiveness of each kind of effort on the feature is called the \emph{effort profile}. The decision maker can observe their performance on the features but cannot observe in which way the agents achieve their scores. \cite{Alon2020} used a similar setting but extended \cite{kleinberg2018}'s model into a multi-agent scenario. Instead of assuming every individual shares the same \emph{effort profile}, they focused on designing optimal algorithms that can work for a group of individuals who may have different effort profiles. Our work is different from theirs with respect to one important aspect: while they assume a feature can be improved in either a causal or non-causal way, our model assumes there are pure causal features and pure correlational features. Causal features (e.g., education level) can be improved only in a `causal' way (such that the value of the target variable will also increase). Correlational features (e.g., whether an applicant wears glasses or not) can be improved only by gaming. If there was a `causal' way to improve the correlational feature, then the firm's willingness to publish the correlational feature might come from the fact that incentivizing individuals to improve will causally increase their work performance. We have shown that even in the case where gaming on the pure correlational feature has no positive effect on productivity, the firm may still want to publish it. Furthermore, none of the papers above have studied how firms may choose between opaque and transparent algorithms. 

Two papers we want to highlight are those of \cite{frankel2019} and \cite{Bonatti2019}. These two economics papers show the decision maker could be ex-ante better off by committing to some ex-post sub-optimal strategies such as down-weighting some relevant features. Our paper is relevant with respect to these papers in the sense that `publishing the algorithms' could also be seen as a way to down-weight the correlational feature. There are critical differences between these papers and ours in terms of mechanisms. In their papers, ex-post sub-optimal behavior such as `under-utilizing' some informative features might be preferred by the decision maker because users may have less incentive to manipulate these features if they anticipate these decisions. The decision maker loses some predictive accuracy from features to true labels, but these features now better represent natural behavior instead of gaming behavior. Consequently, not fully exploiting the information contained in relevant features might be optimal ex-ante. Our paper, however, shows that even if this `down-weighting' behavior has no influence on individuals' gaming behavior, there are still some beneficial effects to the decision maker. In our paper, the purpose of `down-weighting' the correlational feature is not to eliminate gaming behavior, but rather to increase the competitive intensity regarding the causal features for which the \emph{H} type agents hold a cost advantage on.

\vspace{-0.3cm}
\section{Model} \label{Model}
In this section, we discuss a parsimonious model that captures how agents and a firm act under opaque and transparent algorithms. We consider the hiring scenario discussed above with two types of agents: high-talent \emph{H} agents and low-talent \emph{L} agents. For simplicity, we normalize the number of agents to 1 and assume that a $\theta$ portion of them are type \emph{H} and the remaining a $1-\theta$ portion are type \emph{L}. 

Talent level is directly related to job performance and, ideally, the firm would like to hire only high-talent agents. However, the firm cannot directly observe an agent's type until they are hired and work at the firm for a while. Consequently, the firm can only use some observable agent features to help differentiate between these two types of agents. We broadly classify the features into two types: causal features and correlational features. For simplicity, we assume that the firm only uses one causal feature (which is common knowledge to the firm and the agents, for example, education level) and one correlational feature (which is unknown to the agents unless the firm decides to reveal it). Each takes on a discrete value of 0 (low) or 1 (high). In other words, each agent can be characterized by one of four possible combinations (or states) in the feature space:

\begin{itemize}
\item State A (low causal, high correlational);
\item State B (high causal, high correlational);
\item State C (low causal, low correlational); and
\item State D (high causal, low correlational).
\end{itemize}

\noindent The firm's hiring strategy can therefore be represented by four hiring probabilities for the four states. In the remainder of this paper, we will refer to these probabilities as $P_A$, $P_B$, $P_C$, and $P_D$, respectively.

We assume that it is costly for agents to make improvement on the common knowledge causal feature, and that \emph{H} type agents have a cost advantage on this feature. Specifically, we assume that $C_{H}$ (the cost of improving the causal feature for \emph{H} type agents) is less than $C_{L}$ (the cost for \emph{L} type agents). In contrast, the cost for making improvement on the correlational feature is assumed to be the same for \emph{H} type and \emph{L} type agents and is very small (i.e., marginally above zero). It is worth noting that, although the cost of improving any given correlational feature is small, there are many of them, and agents do not know which correlational feature will be used in the algorithm unless the firm decides to reveal it. 

To model the situation where the firm has an incentive to include the correlational feature into its decision making, we further assume that a $\lambda$ portion of \emph{H} type agents and a $1-\lambda$ portion of \emph{L} type agents have value 1 on the firm's chosen correlational feature. Moreover, $\lambda \in [\,0.5,1]$, which, indicates a positive correlation between an agent's value on the correlational feature and their type being \emph{H}. 

The game between the firm and the agents is played as follows. In the first stage, the firm makes a decision on transparency (i.e., opaque or transparent), and this decision is known to all agents. If the firm chooses ``opaque,'' the second stage of the game proceeds as follows. The agents first choose their strategies, and then the firm make its hiring decisions based upon the observed agent strategies. If, on the other hand, the firm chooses ``transparent,'' the second stage of the game proceeds as follows. The firm first announces the correlational feature used in the algorithm together with its hiring strategy (i.e., the probability of hiring an agent at a given state), and then the agents make their decisions. Note that, when choosing the transparent algorithm, the firm reveals \emph{both} the correlational feature it will use and its hiring strategy. This assumption is motivated by common practices. When law makers or the general public ask for ``transparency" of a system, they are requesting the complete knowledge of how that system's algorithms work, including all their input features, their objectives, and their inner-workings. Similarly, when researchers propose algorithmic transparency as a remedy for secret profiling or illegal discrimination, they generally use ``transparency'' to refer to a complete revelation of the mechanism of the algorithm (see, e.g., \citep{Citron} and \citep{theblackboxsociety}).

In this paper, in order to focus on the more interesting and realistic cases, we make the following assumptions regarding the strategy of the agents. 
\begin{enumerate}
\item \emph{In the opaque case, agents will only focus on whether to improve their causal features.} This assumption is motivated by the fact that, while causal features are usually common knowledge between the firm and the agents (e.g., education level plays an important role in hiring decisions), correlational features are less so. In the opaque case, the firm does not reveal which correlational feature it will use in its algorithm to the agents. Consequently, the best that an agent can do is make a random guess. Since there is a large number of potential correlational features and the firm only uses one of them, from the individual agent's perspective, it is not beneficial to improve any of the potential correlational features since the probability of hitting the right one is slight.


\item \emph{In the transparent case, all agents will improve their correlational features.} Once the firm reveals the correlational feature that it will use in its algorithm, the probability that an individual agent to hit the right feature becomes 1. Since the cost of improving the correlational feature is very small, as long as it increases an agent's probability of being hired, they will improve this feature. It is worth noting that assuming all agents will improve the correlational feature used in the algorithm does not cause a loss of generality. There is no loss of generality because, under the scenario where all agents achieve a  ``high'' state on the disclosed correlational feature, this feature completely loses its predictive power and, therefore, will drop out of the prediction algorithm. If it can be shown that the firm can still benefit from making its algorithm transparent in such an extreme case of ``agent gaming,'' it sends a strong message that algorithmic transparency can indeed be economically beneficial.

\end{enumerate}

\subsection{The Firm's Utility}

As previously mentioned, the causal feature (hearafter, we use education level as an example of a causal feature) has a direct influence on the agent's performance, while the correlational feature does not. Thus, an agent's performance is determined by both their type ($T \in \{H, L\}$) and their education level ($Education \in \{0, 1\}$). Here, we are following \cite{weiss1983sorting} and allowing education to not only act as a signal of type but also contribute to the productivity of agents. We assume that the marginal effects of type and education are $\alpha$ and $\beta$, respectively. For convenience, we normalized the performance of an uneducated \emph{L} type agent to 0. The mathematical expression of an agent's performance is thus given by: 
\begin{equation}
W(T, Education) = \alpha \times \mathbbm{1}(T=H) + \beta \times \mathbbm{1}(Education=1).
\label{B1}
\end{equation}

In the opaque case, each agent is characterized by their type $T \in \{H, L\}$ and state (at the end of the game, as defined above) $S \in \{$A, B, C, D$\}$. We can write an agent's performance as a function of their type and final state as follows:
\begin{equation}
W_{S}^{T} = \alpha \times \mathbbm{1}(T=H) + \beta \times (\mathbbm{1}(S=B)+\mathbbm{1}(S=D)).
\label{B2}
\end{equation}

\noindent
In the transparent case, since all agents are ``high'' on the correlational feature, they only differ on the causal feature (i.e., education). This means we can reduce the number of possible states from four to two, state E (low education) and state F (high education). Using these state definitions, we can write an agent's performance as follows:
\begin{equation}
W_{S}^{T} = \alpha \times \mathbbm{1}(T=H) + \beta \times \mathbbm{1}(S=F).
\label{B4}
\end{equation}

Once an agent is hired, their performance will contribute to the firm's payoff, and the firm will pay them reward $R$ (i.e., job compensation). Let $n^T_S$ denote the number of agents of type $T$ whose final states are $S$, and let $n_S = n^H_S + n^L_S$ denote the total number of agents whose final states are $S$. Furthermore, let $\gamma^E_S$ denote the proportion of \emph{H} type agents in state $S$ at the end (E) of the game. The firm's total payoff under hiring strategies (or probabilities) $P = (P_A, P_B, P_C, P_D)$, in the opaque case, or $P = (P_E, P_F)$, in the transparent case, can be mathematically expressed as
\begin{eqnarray}
\Pi_{firm}  & = & \sum_{S} P_S \cdot \left[n^H_S (W^H_S - R) + n^L_S (W^L_S - R)\right] \nonumber \\
& = &  \sum_{S} P_S \cdot n_S \cdot \left[\gamma^E_S (W^H_S - R) + (1-\gamma^E_S) (W^L_S - R)\right]. \label{B3}
\end{eqnarray}

\subsection{The Agents' Utility}

In the opaque case, agents do not know which correlational feature will be used by the firm's algorithm, and therefore, they will only focus on whether to improve on the causal feature (i.e., education). Agents of the same type use the same strategy. Let $u_{T}$ denote the utility of a $T$ type agent, and we have 
\begin{equation}
  u_{H} =
  \begin{cases}
                \lambda P_{B} R+ (1-\lambda) P_{D} R-C_{H}  & \text{if \emph{H} type improves on the causal feature,} \\
                \lambda P_{A} R+ (1-\lambda) P_{C} R & \text{otherwise;} \\
  \end{cases}
\label{B11}
\end{equation}

\noindent

\begin{equation}
  u_{L} =
  \begin{cases}
                (1-\lambda) P_{B} R+ \lambda P_{D} R-C_{L}  & \text{if \emph{L} type improves on the causal feature,} \\
                (1-\lambda) P_{A} R+ \lambda P_{C} R & \text{otherwise.} \\
  \end{cases}
\label{B12}
\end{equation}

In the transparent case, all agents will have a ``high'' value on the correlational feature, and their decision is whether to improve on the causal feature or not. The utility of a $T$ type agent in the transparent case is:
\begin{equation}
  u_{H} =
  \begin{cases}
                 P_{F} R-C_{H}  & \text{if \emph{H} type improves on the causal feature,} \\
                 P_{E} R & \text{otherwise;} \\
  \end{cases}
\label{B13}
\end{equation}

\noindent

\begin{equation}
  u_{L} =
  \begin{cases}
                 P_{F} R-C_{L}  & \text{if \emph{L} type improves on the causal feature,} \\
                 P_{E} R & \text{otherwise.} \\
  \end{cases}
\label{B14}
\end{equation}

\subsection{Additional Parametric Assumptions}

\begin{figure}[htbp]
\centering
\includegraphics[scale=0.55]{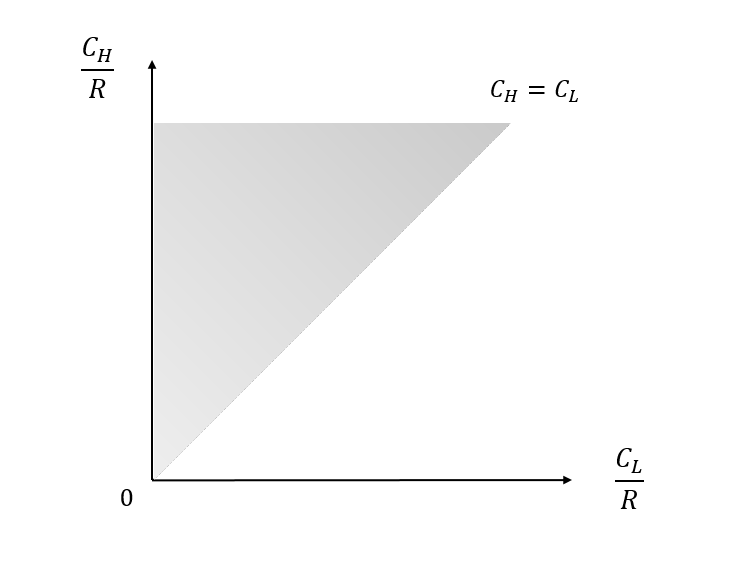}
\caption{The $C_{H}\text{-}C_{L}$ space}
\label{CHCL space}
\end{figure}

In Section \ref{Analysis}, we will solve the game using backward induction. For each combination of $(C_{H}, C_{L})$, we will derive the payoff for the firm in both the opaque and transparent cases. We will then specify the range of values of the parameters where the firm is better or worse off when choosing to be transparent instead of opaque. We will show most of our results in the $C_{H}$-$C_{L}$ space (see Figure \ref{CHCL space}). Since we assume that \emph{H} type agents have a cost advantage on improving the causal feature ($C_{H}<C_{L}$), the region above the diagonal line in Figure \ref{CHCL space} is infeasible.  

We further make the following three assumptions regarding the relationships among different parameters to allow us to focus on non-trivial and more interesting cases.

\begin{assumption}\label{as:1}
  $$0< \beta<R<\alpha.$$
\end{assumption}

Assumption 1 says that the performance of an individual \emph{H} type agent always exceeds the salary $R$ regardless of their education level, whereas the performance of an individual \emph{L} type agent is always smaller than salary $R$. This condition ensures that the firm only wants to hire \emph{H} type agents.

The assumption that $\beta<\alpha$ implies that an \emph{H} type agent with a low level of education is still more productive than an \emph{L} type agent with a high level of education. Furthermore, $\beta<R$ ensures that improving education alone does not guarantee an agent will be hired. If $\beta\geq R$, then the firm will hire anyone with a high level of education, resulting in a trivial equilibrium where everyone receives high education and the firm hires everyone. $\alpha>R$ ensures that the firm will hire an \emph{H} type agent irrespective of their education level. If $\alpha \leq R$, then the firm will once more have no incentive to hire agents without education since their productivity is lower than the salary, leading to an uninteresting equilibrium where everyone receives education.

\begin{assumption}\label{as:2}
    $$\frac{(\theta \lambda+(1-\theta)(1-\lambda))R}{\theta \lambda}<\alpha<\frac{R}{\theta}.$$
\end{assumption}


Recall that $\alpha$ denotes the productivity advantage that \emph{H} type agents hold over \emph{L} type agents. This assumption ensures that $\alpha$ falls in a certain range and does not lead to trivial equilibria. The derivations of the lower bound and the upper bound of $\alpha$ can be found in Appendix \ref{Proof of alpha and beta}. Intuitively, if $\alpha$ is too small, the firm will not have enough incentive to hire anyone even if a large portion of agents are of type $H$. If alpha is too large, the firm will hire everyone even if only a small portion of them are of type $H$. We refer to the lower (upper) bound of $\alpha$ as $\barbelow{\alpha}$($\bar{\alpha}$) hereafter.



\begin{assumption}\label{as:3}
  $$R-\theta \alpha<\beta<R-\frac{\theta(1-\lambda)\alpha}{\theta(1-\lambda)+(1-\theta)\lambda}.$$
\end{assumption}

Assumption \ref{as:3} says that the marginal effect of education on productivity ($\beta$) is in a certain range. This assumption also helps us focus on only non-trivial scenarios. The derivations and interpretations of the lower and upper bound of $\beta$ can be found in Appendix \ref{Proof of alpha and beta}. Intuitively, if $\beta$ is too small, the firm will have little incentive to hire agents with high levels of education. If $\beta$ is too large, then the firm will hire everyone with high levels of education in both the opaque and transparent cases. We refer to the lower (upper) bound of $\beta$ as $\barbelow{\beta}$($\bar{\beta}$) hereafter.

\vspace{-0.3cm}
\section{Analysis}\label{Analysis}


Let $\gamma^B_S$ denote the proportion of \emph{H} type agents in state $S$ at the beginning (B) of the game. Per our discussions in Section \ref{Model}, a $\theta$ portion of agents are of \emph{H} type and the remaining $1-\theta$ portion are of \emph{L} type. Moreover, a $\lambda$ portion of \emph{H} type agents and a $1-\lambda$ portion of \emph{L} type agents have the value 1 for the firm's chosen correlational feature. Therefore,   

\begin{eqnarray}
\gamma^B_{A} & = & \frac{\lambda \theta}{\lambda \theta+(1-\lambda)(1-\theta)}; \\
\gamma^B_{C} & = & \frac{(1-\lambda) \theta}{(1-\lambda) \theta+\lambda(1-\theta)}; \\
\gamma^B_{B} & = & \gamma^B_{D}=0.
\end{eqnarray}

As discussed earlier, for both the opaque and transparent scenarios, we employ the Nash equilibrium as the solution concept. This concept allows us to demonstrate the results in situations where the firm or the agents do not have commitment power or coordination ability as the first mover. Also, in this case, the Nash equilibrium is a more restrictive equilibrium concept than the \emph{Stackelberg equilibrium} when comparing firm payoffs between the opaque and transparent scenarios. The Stackelberg equilibrium would provide the firm a first mover advantage in the transparent case whereas the Nash equilibrium would not. 

\subsection{Opaque Scenario}

Per our discussions in Section \ref{Model}, in the opaque scenario, agents move first and will only decide on whether to improve on the causal feature (i.e., education). We consider equilibrium outcomes where all \emph{H} type agents use the same strategy and all \emph{L} type agents also use the same strategy (see below for discussions on different equilibrium outcomes). 
Given the agents' strategy, based on Equation \ref{B3}, the firm will be indifferent between hiring and not hiring
agents with a final state $S$ if $\gamma^E_S (W^H_S - R) + (1-\gamma^E_S) (W^L_S - R) = 0$, or equivalently,
\begin{equation}
\gamma^E_S = \frac{R - W^L_S}{W^H_S - W^L_S}.
\end{equation}

\noindent
By Equation \ref{B2}, the above fraction equals $\frac{R}{\alpha}$ when $S \in \{$A,C$\}$ and $\frac{R-\beta}{\alpha}$ when $S \in \{$B,D$\}$. Let $\gamma_{th0} = \frac{R}{\alpha}$ and $\gamma_{th1} = \frac{R-\beta}{\alpha}$ (by Assumption \ref{as:1}, we have $0<\gamma_{th1}<\gamma_{th0}<1$). Quantities $\gamma_{th0}$ and $\gamma_{th1}$ are important in the analysis, especially in determining whether a certain outcome (i.e., a joint strategy of the agents and the firm) can be sustained in the equilibrium. 


There are a total of nine different outcomes for agents' strategies. The equilibrium can be sustained with the first five but not the last four. The first five cases are: case 1 (neither \emph{H} type nor \emph{L} type agents improve on education); case 2 (only \emph{H} type agents improve on education); case 3 (both \emph{H} type and \emph{L} type agents improve on education); case 4 (\emph{H} type agents improve on education with some probability, and \emph{L} type agents do not improve on education); and case 5 (\emph{H} type agents improve on education, and \emph{L} type agents improve on education with some probability). 

Aside from these five cases, there are four other cases: case 6 (only \emph{L} type agents improve on education; case 7 (\emph{L} type agents improve on education with some probability but \emph{H} type agents do not improve on education); case 8 (\emph{L} type agents improve on education, but \emph{H} type agents improve on education only with some probability); and case 9 (both \emph{H} type and \emph{L} type agents improve on education with some probability). However, it is not difficult to see that cases 6 through 9 cannot be sustained in the equilibrium. For cases 6 through 8, \emph{L} type agents have a higher value on education than \emph{H} type agents, and the firm will have an incentive to set higher hiring probabilities in states A and C rather than in states B and D. However, under this hiring strategy, \emph{L} type agents will have no incentive to improve on education in the first place. As for case 9, the fact that both \emph{H} type and \emph{L} type agents are using mixed strategies indicates that they are both indifferent between receiving higher education. Since the cost of receiving higher education for \emph{L} type agents is greater than it is for \emph{H} type agents, to compensate for this higher cost \emph{L} type agents must have a higher chance of being hired by the firm than \emph{H} type agents in the equilibrium. However, the firm has no incentive to use such hiring strategy. We conclude that although there are nine possible outcomes for the agents' strategy, only five of them (cases 1 through 5) can be equilibrium outcomes. 

Out of the above five feasible cases, the actual equilibrium strategy of the agents depends on the values of $(C_H, C_L)$. The following lemma summarizes the agents' equilibrium strategy for different values of $(C_H, C_L)$ and the corresponding total payoff for the firm under each equilibrium case. 

\begin{figure}[htbp]
\centering
\includegraphics[scale=0.48]{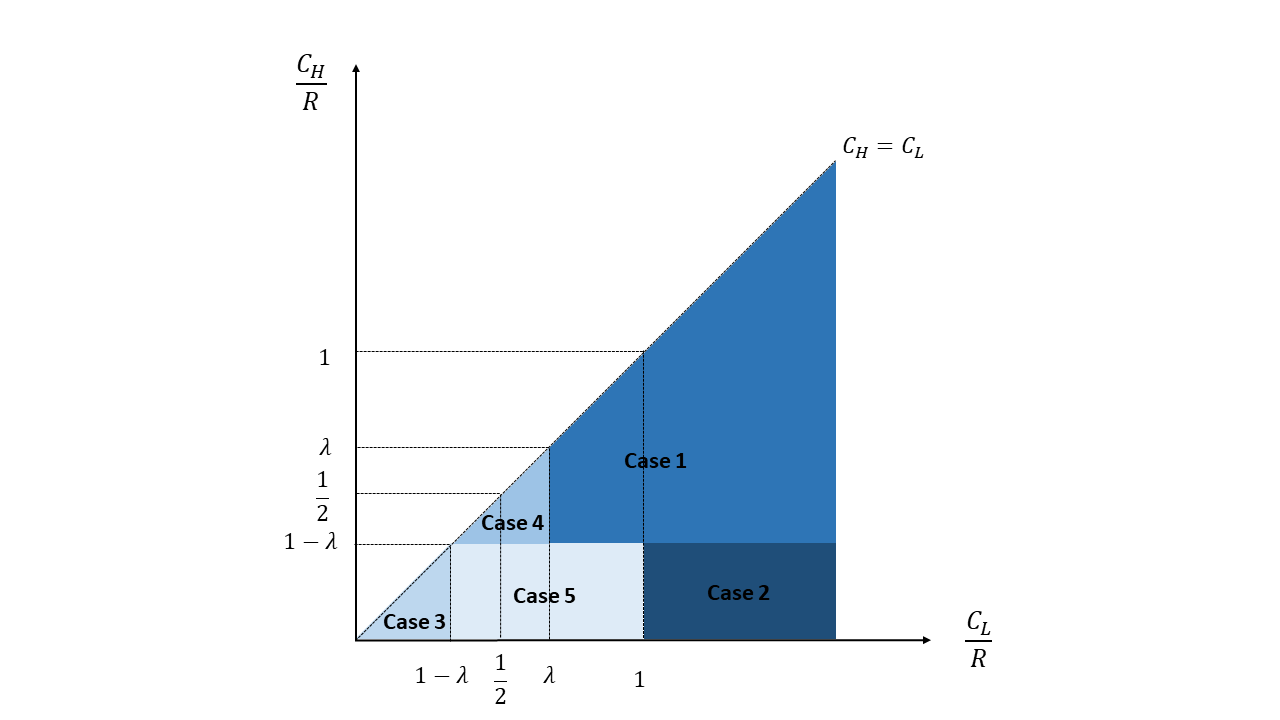}
\caption{Equilibrium outcome in the opaque scenario}
\label{fig:opaque all}
\end{figure}

\begin{lemma}
The equilibrium outcome depends on the values of $(C_{H}, C_L)$, and this dependence is shown in Figure \ref{fig:opaque all}. The corresponding total payoffs for the firm are given by
\begin{align*}
& \Pi_{firm_{O1}}=\lambda \theta \alpha-(\lambda \theta+(1-\lambda)(1-\theta))R \\
& \Pi_{firm_{O2}}=\theta (\alpha+\beta)-\theta R \\
& \Pi_{firm_{O3}}=\lambda \theta (\alpha+\beta)+(1-\lambda)(1-\theta)\beta-(\lambda \theta+(1-\lambda)(1-\theta))R \\
& \Pi_{firm_{O4}}=\theta(\alpha+\beta-R)\left(1-\frac{R(1-\theta)(1-\lambda)}{(\alpha-R)\theta \lambda}\right)\\
& \Pi_{firm_{O5}}=\frac{2\lambda-1}{\lambda}\theta(\alpha+\beta-R),
\end{align*}
\noindent
where $\Pi_{firm_{Oi}}$ denotes the firm's total payoff in case $i$.
\label{lemma1}
\end{lemma}

The proof can be found in Appendix~\ref{Proof of opaque}. In proving Lemma~\ref{lemma1}, we first identify the region in the $C_H$-$C_L$ space in which the strategy of the agents in each of the cases explained above can be sustained as a Nash equilibrium. It turns out that some of these regions overlap (i.e., for a given combination of $(C_H, C_L)$ in the region, there are multiple Nash Equilibria). Since the agents move first in the opaque scenario, they can choose the strategy that gives the highest utility at the end of the game. We focus on the equilibrium outcome that maximizes the utilities for both agent types (i.e., payoff dominant, neither \emph{H} type or \emph{L} type agents have an incentive to deviate from this strategy).

\subsection{Transparent Scenario} 

In the transparent scenario, the firm moves first by announcing both the correlational feature that it wants to use and the probability of hiring for each state (i.e., $P_E$ and $P_F$). Per our discussions in Section \ref{Model}, all agents will improve on this correlational feature and then decide whether to improve on the causal feature (i.e., education). Similar to the opaque scenario, there is a total of nine possible outcomes for the agents' strategy (we use the same numbering of the nine cases as in the opaque scenario). To determine whether a certain market-level outcome (i.e., a joint strategy of the agents and the firm) can be sustained as a Nash equilibrium, we use the fact that the firm is indifferent between hiring and not hiring agents with final state $S$ iff $\gamma^E_S = \frac{R - W^L_S}{W^H_S - W^L_S}$. The fraction on the right side of the equality equals $\gamma_{th0}$ when $S = E$ and $\gamma_{th1}$ when $S = F$. 

BAs with the opaque scenario, cases 6 through 9 cannot be sustained in an equilibrium. 
Similarly, according to Assumption \ref{as:3}, if everyone is at state F, the firm will hire all agents. Under case 5, some \emph{L} type agents are actually at state E, which gives the firm even more incentive to hire agents in state F. However, again, to be a sustainable equilibrium, the mixed strategy outcome in case 5 requires the firm to be indifferent between hiring and not hiring agents from state F. 
Therefore, neither case 4 nor case 5 can be sustained in an equilibrium in the transparent scenario. 
Altogether, this leaves us with only the first three cases as possible equilibrium outcomes. The following lemma summarizes the agents' equilibrium strategy for different values of $(C_H, C_L)$, as well as the corresponding total payoff for the firm under each equilibrium. The proof can be found in Appendix~\ref{Proof of transparent}.

\begin{figure}[htbp]
\centering
\includegraphics[scale=0.48]{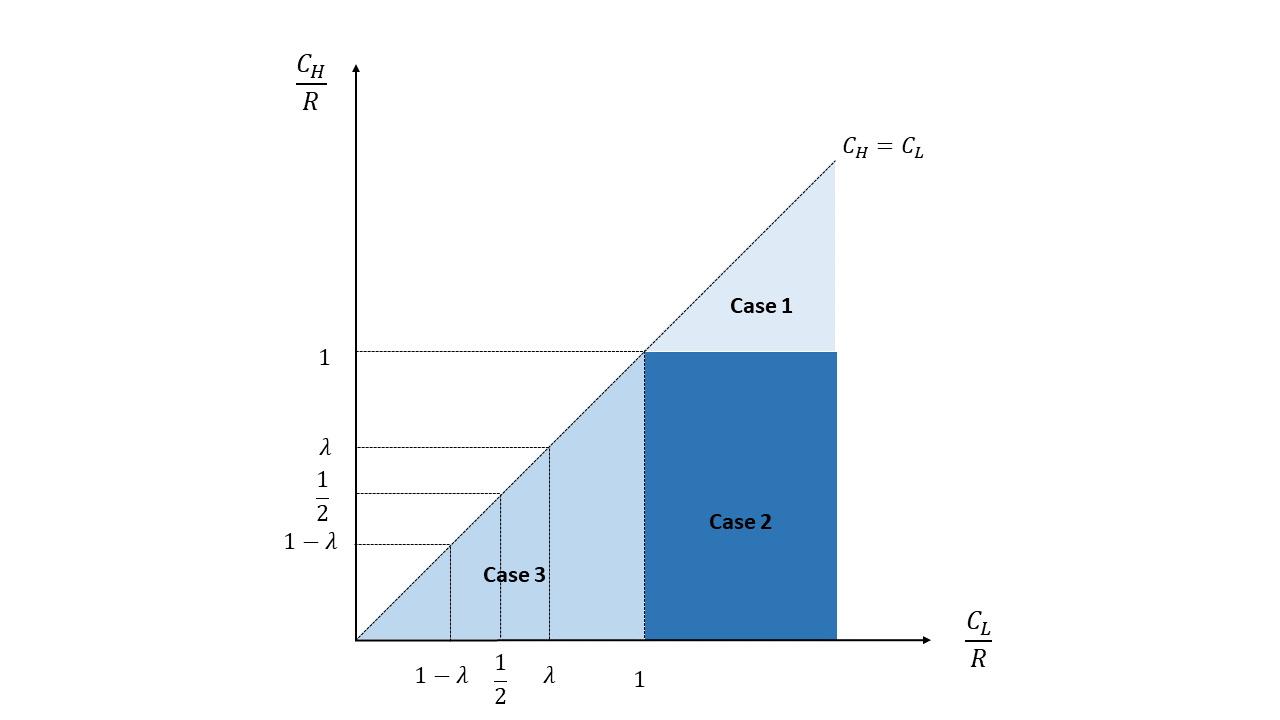}
\caption{Equilibrium outcome in the transparent scenario}
\label{fig:transparent all}
\end{figure}

\begin{lemma}
The equilibrium outcome depends on the values of $(C_{H}, C_L)$, and this dependence is shown in Figure \ref{fig:transparent all}. The corresponding total payoffs for the firm are given by
\begin{align*}
& \Pi_{firm_{T1}}=0 \\
& \Pi_{firm_{T2}}=\theta (\alpha+\beta-R)\\
& \Pi_{firm_{T3}}= \theta (\alpha+\beta)+(1-\theta)\beta-R,
\end{align*}
\noindent
where $\Pi_{firm_{Ti}}$ denotes the firm's total payoff in case $i$.
\label{lemma2}
\end{lemma}

\subsection{The Firm's Decision on Algorithmic Transparency}\label{Transparency-vs-Opacity}

\begin{figure}[htbp]
\centering
\includegraphics[scale=0.48]{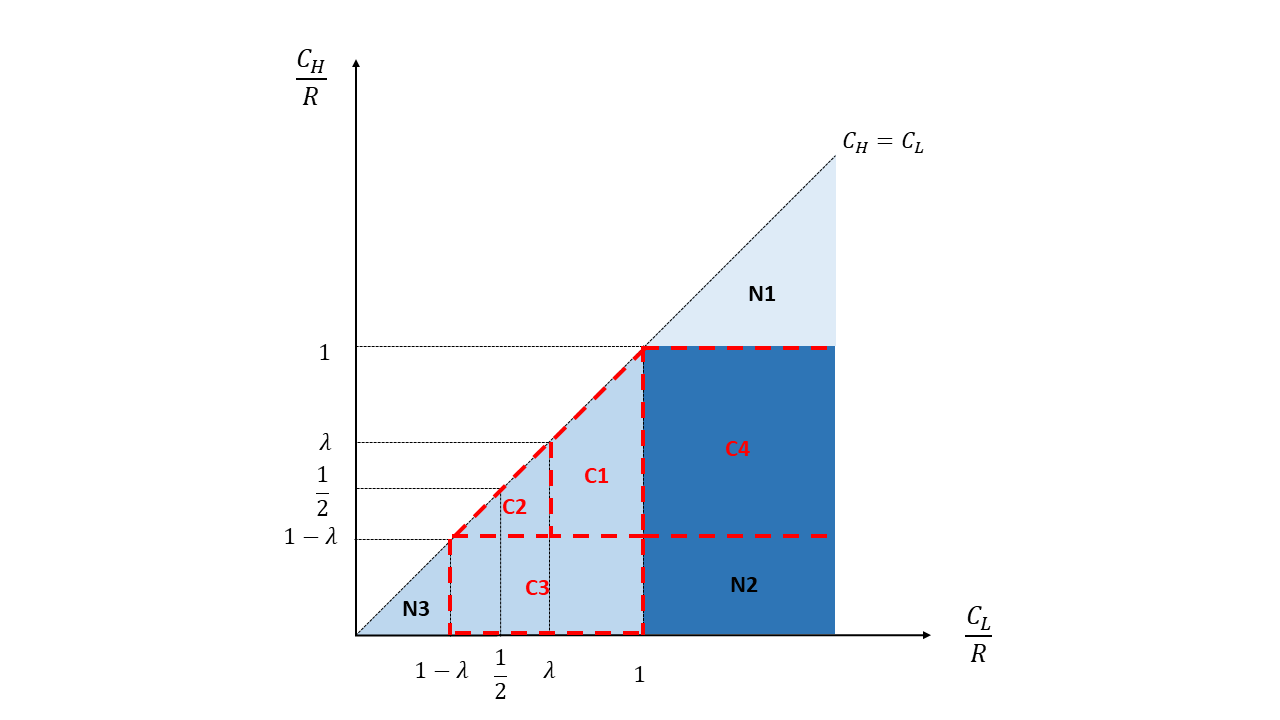}
\caption{The comparison of agents' behavior in the transparent and opaque scenarios}
\label{fig:region change}
\end{figure}

The firm can make a decision on algorithmic transparency by comparing the payoffs in the transparent and opaque scenarios. In this subsection, we will show how the firm is not always worse off when choosing to be transparent. We divide the blue region in Figures \ref{fig:opaque all} and \ref{fig:transparent all} into seven smaller regions: $N1$, $N2$, and $N3$ and $C1$, $C2$, $C3$, and $C4$, respectively (see Figure \ref{fig:region change}).

We first consider regions $N1$ through $N3$. Note that, in these regions, agents apply the same strategy on the causal feature in both the opaque and transparent scenarios. For example, region $N1$ corresponds to case 1 in both the opaque and transparent scenarios, where no agents improve on education. We now discuss the payoff comparison in these regions:
\begin{itemize}
\item In region $N1$, agents play the strategy in case 1 in both the opaque and transparent scenarios.\footnote{It is worth mentioning that although agents' strategies on the causal feature are the same in the opaque and transparent scenarios, the firm's payoff is different because of the existence of the correlational feature.} Mathematically, $\Pi_{firm_{O1}}>\Pi_{firm_{T1}} = 0$. Therefore, the firm will always prefer to be opaque in this region.

\item In region $N2$, agents play the strategy in case 2 in both the opaque and transparent scenarios. Since $\Pi_{firm_{O2}}=\Pi_{firm_{T2}}$, in this region, the firm is indifferent between being opaque or transparent.

\item In region $N3$, agents play the strategy in case 3 in both the opaque and transparent scenarios. We can rewrite $\Pi_{firm_{T3}}$ as follows:
\begin{eqnarray*}
\Pi_{firm_{T3}} & = & \theta(\alpha+\beta)(\lambda+(1-\lambda))+(1-\theta)\beta(\lambda+(1-\lambda))\\
&& - R(\theta\lambda+(1-\theta)(1-\lambda)+(1-\theta)\lambda+(1-\lambda)\theta).
\end{eqnarray*}
Since $\Pi_{firm_{O3}}=\lambda \theta (\alpha+\beta)+(1-\lambda)(1-\theta)\beta-(\lambda \theta+(1-\lambda)(1-\theta))R$, we have:
\begin{eqnarray*}
\Pi_{firm_{T3}}-\Pi_{firm_O3} & = & (1-\lambda)\theta(\alpha+\beta-R)-\lambda(1-\theta)(R-\beta)\\
& = & \theta(1-\lambda)\alpha -(\theta(1-\lambda)+(1-\theta)\lambda)(R-\beta) < 0,
\end{eqnarray*}
where the inequality follows Assumption \ref{as:3}. This means that the firm will always prefer to be opaque in this region. 
\end{itemize}

We conclude that, in regions $N1$ through $N3$, being transparent is never strictly better than being opaque. This is quite intuitive since, in these regions, agents play the same strategy on the causal feature in both the opaque and transparent scenarios. Hence, the firm could only be worse off when revealing its algorithm due to correlational feature's loss of predictive power. Specifically, in regions $N1$ and $N3$, even when the firm chooses to be opaque, the predictive power of the algorithm only comes from the correlational feature because \emph{H} type agents and \emph{L} type agents play the same strategy on the causal feature. This suggests that the firm will incur a significant loss due to the reduction in the algorithm's prediction accuracy when the algorithm is made transparent.

We now discuss the payoff comparison in regions $C1$ through $C4$. It turns out that, in each of these regions, it is possible for the firm to be strictly better off by being transparent instead of being opaque or, in other words, when the value of $\beta$ (the marginal effect of education on an agent's performance) is sufficiently large. In what follows, we first provide a summary of the payoff comparison result in each region, and then discuss the intuition. 
\begin{itemize}
\item In region $C1$, agents' strategies and the firm's payoff change from opaque case 1 to transparent case 3. 
$\Pi_{firm_{T3}} > \Pi_{firm_{O1}}$ iff 
\begin{equation}
\beta > \beta_{1} = \lambda \theta(\alpha-R)-(1-\lambda)(1-\theta)R+R-\theta \alpha.
\label{beta condition 1}
\end{equation}

Thus, in this region, the firm will prefer to be transparent when $\beta > \beta_{1}$.

\item In region $C2$, agents' strategies and the firm's payoff change from opaque case 4 to transparent case 3. $\Pi_{firm_{T3}} > \Pi_{firm_{O4}}$ iff
\begin{equation}
\beta > \beta_{2} = R-\frac{\alpha R(1-\lambda)}{(\alpha-R)\lambda+R(1-\lambda)}.
\label{beta condition 2}
\end{equation}

Thus, in this region, the firm will prefer to be transparent when $\beta > \beta_{2}$.

\item In region $C3$, agents' strategies and the firm's payoff change from opaque case 5 to transparent case 3. $\Pi_{firm_{T3}} > \Pi_{firm_{O5}}$ iff
\begin{equation}
\beta > \beta_{3} = \frac{\lambda\theta\alpha-\theta\alpha-2\lambda\theta R+\theta R+\lambda R}{\lambda-2\theta\lambda+\theta}.
\label{beta condition 3}
\end{equation}

In this region, the firm will prefer to be transparent when $\beta > \beta_{3}$. It is interesting to note, however, that $\beta_3$ equals $\bar{\beta}$ 
defined in Assumption \ref{as:3} (see below for an explanation of why this is the case). Since the value of $\beta$ cannot exceed $\bar{\beta}$ (according to Assumption \ref{as:3}), this means that the firm will never prefer to be transparent in this region.

\item In region $C4$, agents' strategies and the firm's payoff change from opaque case 1 to transparent case 2. $\Pi_{firm_{T2}} > \Pi_{firm_{O1}}$ regardless of $\beta$. Thus, in this region, the firm will always prefer to be transparent.
\end{itemize}

According to Assumption \ref{as:2}, the value of $\alpha$ falls in the following interval: 
$$\frac{(\theta \lambda+(1-\theta)(1-\lambda))R}{\theta \lambda}<\alpha<\frac{R}{\theta}.$$ 
Within the above interval, $\beta_{1}$ and $\beta_{3}$ are decreasing in $\alpha$, and $\beta_{2}$ is increasing in $\alpha$. Moreover, $\beta_{1}=\beta_{2}$ when $\alpha=\barbelow{\alpha}$ and $\beta_{2}=\beta_{3}$ when $\alpha=\bar{\alpha}$ ($\barbelow{\alpha}$ and $\bar{\alpha}$ are defined in Assumption 2). Thus, we have $\beta_{1}<\beta_{2}<\beta_{3}$. To understand why we have increasing thresholds for $\beta$ as we move from regions $C1$ to $C3$ and why there is no threshold for $\beta$ in region $C4$, we must look at how the firm's decision to be transparent changes the agents' strategies in different regions. To facilitate our discussions, we first define the 
concept of ``degree of separation." 
Suppose that there are $n_{H0}$ \emph{H} type agents and $n_{L0}$ \emph{L} type agents who do not improve on education and $n_{H1}$ \emph{H} type agent and  $n_{L1}$ \emph{L} type agents who improve on education. 
We define the degree of separation ($Dos$) between \emph{H} type and \emph{L} type agents as follows: $$Dos = 1-\frac{\min(n_{H0},n_{L0})+\min(n_{H1},n_{L1})}{n_{H0}+n_{L0}+n_{H1}+n_{L1}}.$$

\noindent
Note that, if all agents improve on education or no one improves on education, then $Dos$ reaches its minimum value: $Dos_{min} = \max(\theta,1-\theta)$. If all \emph{H} type agents improve on education and no \emph{L} type agents improve on education, then $Dos$ reaches its maximum value: $Dos_{max} = 1$. If either \emph{H} type or \emph{L} type agents use a mixed strategy, the value of $Dos$ is somewhere in between. 
The key observation here is that, the higher the value of $Dos$, the easier it is for the firm to differentiate \emph{H} type agents from \emph{L} type agents using the causal feature. 

We now discuss how the firm's decision to be transparent changes agents' strategies in regions $C1$ through $C4$. 
Note that transparency intensifies agents' competition on the causal feature and that this intensified competition has the following two effects. 
\begin{enumerate}
    \item \emph{The degree of separation between \emph{H} type and \emph{L} type agents on their causal feature changes}. As an illustration, consider region $C4$. In this region, when the firm switches from being opaque to being transparent, agents' strategies also switch from opaque case 1 to transparent case 2. Under opaque case 1, neither \emph{H} type nor \emph{L} type agents improve on education, whereas under transparent case 2, only \emph{H} type agents improve on education. This means that \emph{H} type agents are now more separated from \emph{L} type agents on the causal feature (i.e., there is a higher degree of separation). Similarly, it can also be verified that, in regions $C2$ and $C3$, agents become less separated and, in region $C1$, there is no change in the degree of separation.
    
    \item \emph{Agents' average value on the causal feature becomes higher and their work performance increases (according to Equation \ref{B2}).} To see this, consider region $C1$. In this region, when the firm switches from being opaque to being transparent, agents' strategies also switch from opaque case 1 to transparent case 3. Although the change in agents' strategies does not affect the degree of separation, since both types of agents improve on education, the average level of education for both agent types increases. Similarly, in region $C2$, the average level of education and, thus, the performance level of both types of agents increase. In region $C3$, only the average performance of \emph{L} type agents increases, whereas in region $C4$, only the average performance of \emph{H} type agents increases. We can see that, in regions $C1$ through $C4$, the overall average agents' performance always increases when the firm switches from being opaque to being transparent. 
    
\end{enumerate}

Since the firm always loses useful information from the correlational feature that helps it differentiate between the two types of agents when switching from the opaque to the transparent algorithm, the firm's decision on algorithmic transparency will depend on whether the above two effects (i.e., the change in the degree of separation and the increase in the agents' average performance) can offset the negative effect of information lost on the correlational feature. In region $C1$, even though the degree of separation does not change, both \emph{H} type and \emph{L} type agents improve on education, and the firm can benefit from the increase in the average agents' performance. Whether this benefit offsets the negative effect of information loss on the correlational feature depends on the value of $\beta$. If $\beta$ is large enough (i.e., $\beta > \beta_{1}$), then being transparent is preferred over being opaque. In region $C2$, the degree of separation decreases as the agents' distribution changes from partial separation to pooling. However, the average performance of \emph{H} type and \emph{L} type agents increases, which suggests that, if $\beta$ is large enough, the firm can still be better off being transparent. The conditions on $\beta$ in this case is stricter than in region $C1$ (i.e., $\beta > \beta_2 > \beta_1$). This is because the marginal effect of education must now be large enough to offset not only the previously mentioned negative effect of the loss of information on the correlational feature, but also the worse degree of separation on the causal feature.

In region $C3$, the firm's decision to be transparent affects fewer agents compared to in region $C2$. Switching from opaque to transparent incentivizes all \emph{L} type agents and some \emph{H} type agents in region $C2$ to improve on education but only incentivizes some \emph{L} type agents to improve on education in region $C3$. Since fewer agents are affected by the firm's switching from opaque to transparent in region $C3$ compared with region $C2$ and since the increase in the agents' average performance is proportional to the number of agents being affected, a larger $\beta$ is needed in region 3 to achieve the same level of average performance found in region $C2$. This is why $\beta_3 > \beta_2$. 

To see why $\beta_3 = \bar{\beta}$ (as defined in Assumption \ref{as:3}), note that all \emph{H} type agents have already improved on education in the opaque case, and, only some \emph{L} type agents will switch from not improving to improving on education when the algorithm is made transparent. According to Assumption \ref{as:1}, the individual productivity of \emph{L} type agents cannot exceed $R$, which means that the firm will not benefit from hiring extra \emph{L} type agents even if their education levels are high. In fact, the portion of \emph{L} type agents who improve on education also depends on $\beta$ (see Equation \ref{eq:pL} in the Appendix). When the value of $\beta$ is close to $\bar{\beta}$, nearly all \emph{L} type agents have already chosen to improve on education in the opaque case. Therefore, switching from opaque to transparent makes a small difference in terms of agents' average performance. However, the firm will be indifferent between being opaque and transparent only when $\beta$ reaches $\bar{\beta}$. This is why $\beta_3$ equals $\bar{\beta}$. 

In region $C4$, the degree of separation increases when the firm switches from being opaque to being transparent. In fact, the agents' distribution changes from pooling to perfect separation. This effect in itself is sufficient to offset the negative effect of information lost on the correlational feature. This is the reason why, in this region, the firm prefers to be transparent 
regardless of the value of $\beta$. 

The following theorem summarizes our findings about the firm's decision on transparency:

\begin{theorem}

In regions $N1$ through $N3$, being transparent is never strictly better than being opaque. In regions $C1$ through $C4$, depending on the value of $\beta$, the firm may prefer being transparent to being opaque. Specifically, in region $C1$, the firm will prefer to be transparent if $\beta>\beta_{1}$; in region $C2$, the firm will prefer to be transparent if $\beta>\beta_2$; in region $C3$, the firm will never prefer to be transparent; and, in region $C4$, the firm will prefer to be transparent regardless of the value of $\beta$.
\label{theorem1}
\end{theorem}

\subsection{The Effect of the Predictive Power of the Correlational Feature ($\lambda$) and the Fraction of High--Talent Agents ($\theta$)}\label{Comparatives}
We have shown that the firm will be strictly better off by making the algorithm transparent if the ($C_H,C_L$) pair lands in region $C4$ or in regions $C1$ or $C2$ with some additional conditions on $\beta$. In region $C4$, the transparent algorithm is preferred regardless of $\beta$, $\lambda$, and $\theta$ since removing the correlational feature changes agents' behavior on the causal feature from pooling to full separation and full separation is the case in which the firm receives maximum profit. In regions $C1$ and $C2$, the main driving force that makes the transparent algorithm preferable is the agents' increased average performance on the causal feature. As long as $\beta$ exceeds the threshold $\beta_1$($\beta_2$), the increased performance on the causal feature will have a large enough positive effect on work performance to make the firm better off. We now examine how the thresholds on $\beta$ changes when $\lambda$ or $\theta$ changes. In other words, we are going to look at when the correlational feature is more informative (a larger $\lambda$) or when there is a larger portion of $H$ type agents in the population (a larger $\theta$), will the condition on $\beta$ to make the transparent algorithm be preferred becomes stricter or milder. 

Take the expressions of $\beta_1$ and $\beta_2$ in Equations \ref{beta condition 1} and \ref{beta condition 2}, and take derivatives with respect to $\lambda$. 

\begin{equation}
    \frac{\partial \beta_1}{\partial \lambda}=-2\theta R + \theta \alpha +R.
\label{beta1 lambda}
\end{equation}

\begin{equation}
    \frac{\partial \beta_2}{\partial \lambda}= \frac{\alpha R(\alpha-R)}{(2\lambda R-R-\alpha \lambda)^2}.
\label{beta2 lambda}
\end{equation}

Both of these derivatives are greater than 0 in the parameter ranges we are considering. This means that, within each region, as $\lambda$ becomes larger, a higher value of $\beta$ is needed to make the transparent algorithm preferable. This is because a larger $\lambda$ value entails more information is contained in the correlational feature, so a higher causal effect is required to offset the loss of information from the correlational feature. However, the effect of $\lambda$ on algorithmic transparency is not this straightforward since, apart from the equilibrium payoff, $\lambda$ can also determine what kind of equilibrium can be sustained given a ($C_H$,$C_L$) pair (i.e., the regions' shapes in Figure \ref{fig:region change} will change as $\lambda$ changes). Consider the region just beneath the dividing line of regions $N2$ and $C4$: as $\lambda$ increases, it changes from belonging to region $N2$ to belonging to $C4$. This means the firm will prefer the opaque algorithm when facing a small $\lambda$ but prefer a transparent algorithm when faced with a large $\lambda$. Although it appears counterintuitive, it can be explained as follows. When $\lambda$ is small, making the algorithm transparent is not effective enough to change agents' behavior on the causal feature. However, when $\lambda$ is large, agents' behavior on the causal feature will change drastically, and the firm is able to hire agents who are on average more productive under the transparent scenario than the opaque scenario.

The following proposition summarizes our findings about how $\lambda$ affects the firm's decision on transparency:

\newtheorem{prop}{Proposition}
\begin{prop}
An increase in $\lambda$ has the following effects on the firm's decision on transparency: 

\begin{enumerate}
\item The area of regions $C1$, $C2$, and $C4$ increases, which means a transparent algorithm is preferred under more ($C_H$,$C_L$) value pairs. 
\item Within regions $C1$ and $C2$, the conditions on $\beta$ to make a transparent algorithm more preferred become stricter. 
\end{enumerate}
\label{prop:comparatives_lambda}
\end{prop}

\noindent
Take the expressions of $\beta_1$ and $\beta_2$ in Equations \ref{beta condition 1} and \ref{beta condition 2}, and take the derivative with respect to $\theta$. 

\begin{equation}
    \frac{\partial \beta_1}{\partial \theta}=-2\lambda R + \lambda \alpha +R-\alpha.
\label{beta1 theta}
\end{equation}

\begin{equation}
    \frac{\partial \beta_2}{\partial \theta}=0.
\label{beta2 theta}
\end{equation}

It can be shown that $\frac{\partial \beta_1}{\partial \theta}$ is smaller than 0 in the parameter ranges we are considering. This means that, in region $C1$, as the proportion of $H$ type agents increases, the conditions on $\beta$ to make algorithmic transparency more preferred become milder. This is because the degree of separation on the causal feature in region $C1$ does not change (from pooling at 0 to pooling at 1). The only negative effect of algorithmic transparency is the loss of the correlational feature. When $\theta$ is high, losing the correlational feature is less harmful for the firm (less $L$ type agents are mistakenly hired when the correlational feature is lost). Thus, a smaller $\beta$ is needed to offset this negative effect. In region $C2$, however, $\theta$ has no influence on the firm's decision on algorithmic transparency. This is because in region $C2$, there are two negative effects of algorithmic transparency, the loss of the correlational feature and a smaller degree of separation on the causal feature. A higher $\theta$ will mitigate the first effect (similar reason as before) and amplify the second (more $H$ type agents will be left out). Overall, $\theta$ does not affect the value of $\beta$ needed for making the transparent algorithm preferable.

The following proposition summarizes our findings about how $\theta$ affects the firm's decision on transparency:

\begin{prop}
In region $C1$, a higher $\theta$ will increase the firm's incentive to make the algorithm transparent. In other regions, $\theta$ has no impact on the firm's decision on algorithmic transparency.
\label{prop:comparatives theta}
\end{prop}

\subsection{Agents' Welfare}

In Subsections \ref{Transparency-vs-Opacity} and \ref{Comparatives}, we have specified conditions under which a transparent algorithm will yield a strictly higher payoff to the firm. We will next investigate the impact of algorithmic transparency on the agents' welfare. 

The agents' payoff in the equilibrium is summarized in the following lemma.

\begin{lemma}
For each equilibrium outcome shown in Figure \ref{fig:opaque all} (for the opaque scenario) and Figure \ref{fig:transparent all} (for the transparent scenario), the corresponding total payoffs for agents are given by 
\begin{align*}
& \Pi_{agents_{O1}}=(\lambda \theta+(1-\lambda)(1-\theta))R \\
& \Pi_{agents_{O2}}=\theta (R-C_H) \\
& \Pi_{agents_{O3}}=(\lambda \theta+(1-\lambda)(1-\theta))R-C_H\theta -C_L(1-\theta) \\
& \Pi_{agents_{O4}}=\frac{(\lambda \theta+(1-\lambda)(1-\theta))(R-C_H)}{\lambda}\\
& \Pi_{agents_{O5}}=\frac{(2R\lambda-R-C_H\lambda-C_L\lambda+C_L)\theta}{\lambda}\\
& \\
& \Pi_{agents_{T1}}=0 \\
& \Pi_{agents_{T2}}=\theta (R-C_H)\\
& \Pi_{agents_{T3}}= R-C_H\theta-C_L(1-\theta),
\end{align*}
\noindent
where $\Pi_{agents_{Oi}}$ denotes the agents' total payoff in case $i$ of the opaque scenario and $\Pi_{agents_{Ti}}$ denotes the agents' total payoff in case $i$ of the transparent scenario.   

\label{lemma:agents welfare}
\end{lemma}

As previously discussed, Figure \ref{fig:region change} shows how agents' behavior changes on the causal feature when the algorithm is made transparent. First, consider the three regions where the agents' behavior on the causal feature does not change (regions $N1$, $N2$, and $N3$). In directly comparing the agents' payoff in the equilibrium, we have the following.

\begin{itemize}
\item In region $N1$, agents play the strategies in case 1 in both the opaque and transparent scenarios. $\Pi_{agents_{O1}}>\Pi_{agents_{T1}}$. Therefore, agents will receive a higher payoff under opaque algorithm in this region.

\item In region $N2$, agents play the strategies in case 2 in both the opaque and transparent scenarios. Since $\Pi_{agents_{O2}}=\Pi_{agents_{T2}}$, in this region, agents are indifferent to whether the algorithm is opaque or transparent.

\item In region $N3$, agents play the strategies in case 3 in both the opaque and transparent scenarios.
$\Pi_{agents_{O3}}<\Pi_{agents_{T3}}$. Therefore, agents will receive a higher payoff under transparent algorithm in this region.
\end{itemize}

In regions $N1$, $N2$, and $N3$, agents' behavior on the causal feature does not change. Since improving on the correlational feature is assumed to be costless, agents' total cost stays the same as the algorithm becomes transparent. The only thing that varies is the benefit they can obtain under the firm's strategy. In region $N1$, more agents will be hired under opaque algorithm (a $\lambda$ portion of $H$ type agents and a $1-\lambda$ portion of $L$ type agents are hired under the opaque algorithm but no one will be hired under the transparent algorithm). In region $N3$, more agents will be hired under the transparent algorithm (a $\lambda$ portion of $H$ type agents and a $1-\lambda$ portion of $L$ type agents are hired under the opaque algorithm and everyone will be hired under the transparent algorithm). In region $N2$, the same number of agents will be hired regardless of whether the algorithm is opaque or transparent (only $H$ type agents will be hired). 

Furthermore, consider the four regions where agents' behavior on the causal feature changes (regions $C1$, $C2$, $C3$, and $C4$). By directly comparing the agents' payoffs in the equilibrium, we obtain the following. 

\begin{itemize}
\item In region $C1$, agents' strategies and the firm's payoff change from opaque case 1 to transparent case 3. $\Pi_{agents_{T3}} \leq \Pi_{agents_{O1}}$ iff 
\begin{equation}
C_H\theta+C_L(1-\theta) \geq (1-\lambda\theta-(1-\lambda)(1-\theta))R.
\label{agents condition 1}
\end{equation}

The smallest possible value for LHS is reached when a ($C_H$,$C_L$) pair lands at the lower left corner in region $C1$, in other words, when $C_H=(1-\lambda)R$ and $C_L=\lambda R$. It can further be shown that this smallest value equals the RHS. Thus, Equation \ref{agents condition 1} is satisfied for any ($C_H$,$C_L$) value in region $C1$. In this region, the transparent algorithm will give agents a lower payoff compared with the opaque algorithm.

\item In region $C2$, agents' strategies and the firm's payoff change from opaque case 4 to transparent case 3. $\Pi_{agents_{T3}} \geq \Pi_{agents_{O4}}$ iff
\begin{equation}
\frac{1-\lambda}{\lambda}C_H-C_L \geq (\frac{1}{\lambda}-2)R.
\label{agents condition 2}
\end{equation}

The smallest possible value for LHS is reached when a ($C_H$,$C_L$) pair lands at the lower right corner in region $C2$, in other words, when $C_H=(1-\lambda)R$ and $C_L=\lambda R$. It can further be shown that this smallest value equals the RHS. Thus, Equation \ref{agents condition 2} is satisfied for any ($C_H$,$C_L$) value in region $C2$. In this region, the transparent algorithm will give agents a higher payoff compared with the opaque algorithm.

\item In region $C3$, agents' strategies and the firm's payoff change from opaque case 5 to transparent case 3. $\Pi_{agents_{T3}} \geq \Pi_{agents_{O5}}$ iff
\begin{equation}
(2\theta-\frac{\theta}{\lambda}-1)C_L \geq (2\theta-\frac{\theta}{\lambda}-1)R. 
\label{agents condition 3}
\end{equation}

Since $2\theta-\frac{\theta}{\lambda}-1 \leq 0$, and $C_L \leq R$ in region $C3$, Equation \ref{agents condition 2} is satisfied for any ($C_H$,$C_L$) value in region $C3$. In this region, the transparent algorithm will give agents a higher payoff compared with the opaque algorithm.

\item In region $C4$, agents' strategies and the firm's payoff change from opaque case 1 to transparent case 2. $\Pi_{agents_{T2}} \leq \Pi_{agents_{O1}}$ iff

\begin{equation}
 \theta(R-C_H) \leq (\lambda\theta+(1-\lambda)(1-\theta))R.
\label{agents condition 4}
\end{equation}

The largest possible value for LHS is reached when a ($C_H$,$C_L$) pair lands at the lower bound of region $C4$, in other words, when $C_H=(1-\lambda)R$. It can be shown that this largest value is smaller than the RHS. Thus, Equation \ref{agents condition 4} is satisfied for any ($C_H$,$C_L$) value in region $C4$. In this region, the transparent algorithm will give agents a lower payoff compared with the opaque algorithm.

The following theorem summarizes our findings regarding the agents' welfare under opaque and transparent algorithms. 

\end{itemize}

\begin{figure}[htbp]
\centering
\includegraphics[scale=0.45]{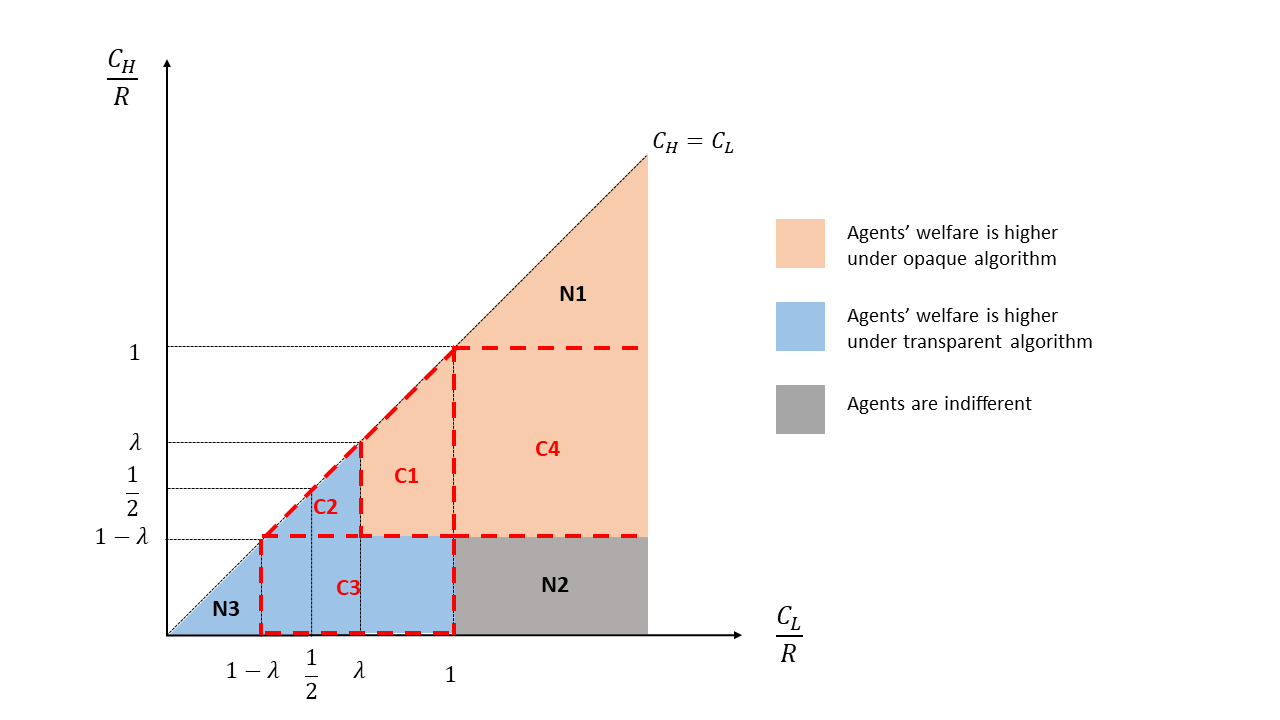}
\caption{The comparison of agents' welfare in transparent and opaque scenarios}
\label{fig:agents' welfare}
\end{figure}

\begin{theorem} Whether the agents are better off under opaque or transparent algorithm depends on the values of ($C_H$,$C_L$), and this dependence is shown in Figure \ref{fig:agents' welfare}. 
In regions $N3$, $C2$ and $C3$, agents' welfare is higher under transparent algorithm. In region $N1$, $C1$ and $C4$, agents' welfare is higher under opaque algorithm. In region $N2$, agents' welfare is not affected by algorithmic transparency.
\label{theorem:agents}
\end{theorem}

The intuition behind Theorem \ref{theorem:agents} can be explained as follows. All agents will prefer an opaque algorithm when $C_H$ and $C_L$ are large. They will prefer a transparent algorithm when $C_H$ and $C_L$ are small, and will be indifferent when $C_L$ is large but $C_H$ is small. Making the algorithm transparent may force agents to invest in the causal feature, but they can also attain a certain benefit at the same time (higher chance of being hired). The cost of this investment increases as $C_H$ and $C_L$ increase, but the benefit does not vary with $C_H$ and $C_L$. Consequently, agents will be worse (better) off under transparent algorithm if $C_H$ and $C_L$ are large (small). 

Comparing Theorem \ref{theorem1} with Theorem \ref{theorem:agents}, we find that the firm and agents' interests conflict in some regions. For example, in regions $N3$ and $C3$, a transparent algorithm will give agents a higher payoff, but the firm prefers an opaque algorithm. In region $C4$, an opaque algorithm will give agents a higher payoff but the firm prefers a transparent algorithm. In regions $C1$ and $C2$, whether their interests conflict depends on the value of $\beta$. 

In region $C2$, if condition $\beta > \beta_2$ is satisfied, both the firm and agents would prefer a transparent algorithm over an opaque algorithm. In other words, making the algorithm transparent will Pareto dominate keeping the algorithm opaque.

\section{Conclusion}\label{Conclusion}
\subsection{Summary of Results}

In this paper, we studied how firm and agent welfare is affected by algorithmic transparency. We allowed the agents to be strategic such that they can invest in their causal and correlational features to increase their chances of being hired in response to a firm's algorithm. We also investigated how the predictive power of the correlational feature, the market composition in terms of the fraction of \emph{H} type agents, and the impact of the causal feature on agent productivity affect the firm's decision to make their algorithm transparent or opaque. 

As a first result, we identified a broad set of conditions under which the firm would be better off with algorithmic transparency rather than opacity. There are two scenarios where a firm would prefer algorithmic transparency. The first is when the $H$ type agents have a significant cost advantage on causal feature. The second is when the $H$ type agents have only a moderate cost advantage on the causal feature, the cost itself is moderate and the causal feature has a significant impact on the agents' productivity. The key intuition behind these conditions is that disclosing the correlational feature will intensify agents' competition on the causal feature, which may either help the firm separate different types of agents more easily or increase agents' average work performance (or both). This finding is important because concerns about agents ``gaming the system'' have been the main reason why firms are reluctant to make their algorithms transparent despite increasing calls for algorithmic transparency.  

Our second result is that the agents may not always be better off under algorithmic transparency. We found that, when the firm prefers algorithmic transparency, the agents prefer opacity and vice versa. Only when specific conditions related to the cost of the causal feature (scenario 2 above) are satisfied will the firm and agents both prefer algorithmic transparency. 
 
Our third result is that, when the correlational feature has high predictive power in the opaque scenario, the firm could be better off making the algorithm transparent. When the predictive power of the correlational feature is nullified due to gaming under algorithmic transparency, the only way agents can signal their type is by investing in the costly causal feature which makes the firm better off.
 
Our final result is that, when the fraction of $H$ type agents on the market is high, the firm would be better off by making its algorithm transparent. Even when the causal feature is unable to separate the two types of agents, the firm is at a much lower risk of hiring an $L$ type agent if they are only a few of them on the market.
 
\subsection{Implications for Managers}
Our paper shows that managers using machine learning models for decision making could be better off by making their algorithms transparent. Algorithmic transparency does not always mean a loss of predictive power. In some cases, it can in fact lead to greater predictive power. In other cases, while it may reduce predictive power, it may still make them better off by improving the desirability of the whole market. Our results are particularly promising for managers, as they are now facing growing calls to make their algorithms transparent.

We identified a set of conditions where managers should prefer algorithmic transparency. There are three factors that managers should consider: (a) access to a good causal feature; (b) the predictive power of the correlational features; (c) the market composition in terms of the fraction of $H$ type agents.

We provide some guidance on what make a good causal feature here. The causal feature serves two purposes: (a) a signaling purpose – $H$ type agents have a cost advantage on this feature, and thus, it can help separate $H$ type agents from $L$ type agents; and (b) a human capital purpose - the feature itself contributes to the productivity of the agents. To serve the signaling purpose well, the identified causal feature should be neither too costly nor too cheap to improve. If it is too costly, no one will improve on it. By contrast, if it is too cheap, everyone will improve on it. Furthermore, in terms of the human capital purpose, the higher the feature's impact on productivity, the better it is. Even when this causal feature is unable to completely separate the $H$ type agents from $L$ type agents, if it is moderately costly and contributes to productivity, the firm could still be better off. Typically algorithm designers are not focused on causality or identifying causal features. Our results indicate that they should. The recent stream of research in computer science that is focused on causal inference in machine learning models bodes well for them in this regard \citep{pearl2000models}. 

The second factor that managers should consider is the predictive power of the correlated features. Intuitively, managers may think of keeping their algorithms opaque when the correlational features provide significant predictive power. Our results show that this thinking is incorrect. We show that firms are more likely to be better off by making their algorithms transparent when correlational features provide significant predictive power in the opaque counterpart. Though incorrectly, managers would be particularly concerned about making the algorithms transparent when the marginal effect of the causal feature in separating the $H$ type agents from $L$ type agents is small in the presence of correlational features in an opaque algorithm. However, they should realize that the effect of the causal feature is suppressed largely due to the strategic behavior of the agents when the algorithm is opaque.  All agents, but more importantly, $H$ type agents, underinvest in the causal feature when they know the correlational feature can separate them from $L$ type agents: the higher the predictive power of the correlational feature, the lower the incentive of the agents to invest in the causal feature. When the algorithm is made transparent, in the new equilibrium, the agents have a significantly higher incentive to invest in the causal feature, making the firm better off.    

To identify how much predictive power comes from each feature in a machine learning model is not a trivial task. However, recent research on algorithm explainability and interpretability \citep{arrieta2020explainable} has started to provide approaches that can help identify what features are important in a machine learning model. Managers should use these approaches to identify what causal and correlational features matter in their machine learning model.

The third factor managers should consider is the market composition in terms of the fraction of $H$ type agents. If managers find that the causal feature is unable to separate $H$ type agents from $L$ type agents, they may still be better off with algorithmic transparency if the cost of improving the causal feature is moderate and the market is composed of more $H$ type agents.

Overall, our results suggest that managers should not view manipulation by agents as bad. Rather, they should embrace it and use algorithmic transparency as a lever for motivating agents to invest in more desirable actions.

\subsection{Implications for Public Policy}
There are two key arguments typically put forth in support of algorithmic transparency. First, making algorithms transparent could highlight any hidden biases in algorithms and make them accountable \citep{lambrecht2019algorithmic, cowgill2020algorithmic}. Second, the users who are affected by an algorithm's decision making have the right to see which factors affect decisions made about them. Our paper has implications related to this second argument. 

Recent legislation like the General Data Protection Regulation has afforded individuals a right to explanation under which firms have to provide an explanation regarding how a decision has been made by their algorithms \citep{goodman2017european}. Our results show that such a regulation may not improve consumer welfare. When agents know which features in an algorithm affect important decisions about them, they can improve those features. Consequently, algorithmic transparency is generally viewed as helping the agents at the cost of the firm. However, our results show that the agents may not benefit from algorithmic transparency as expected. When algorithms are opaque, they derive their accuracy from both causal and correlational features. Therefore, $H$ type agents do not need to invest in the costly causal feature to separate themselves from $L$ type agents. In the transparent scenarios, in many cases, the agents have to invest in the costly causal feature to achieve similar or even less separation with no change in their wage. 

\subsection{Future Research Directions}
Firms have typically kept their algorithms opaque to protect them from gaming by agents. In this study, we show that this strategy may not be the best, and the firm could be better off by making its algorithm transparent in the presence of strategic users. However, there are two other reasons as to why firms may still not want to make their algorithm transparent -- interpretability \citep{kroll, brauneis} and privacy \citep{Price, brauneis}. With access to big data and large computational power, machine learning models have become complicated to the extent that they are rendered uninterpretable. While recent research has made advances in developing interpretable machine learning models \citep{arrieta2020explainable}, \cite{bertsimas2019price} show that model interpretability comes at a cost of accuracy. As a result, when considering the issue of algorithmic transparency, it may be interesting to consider the tradeoff between interpretability and accuracy. Similarly, when a firm makes its algorithm transparent, this can lead to privacy concerns. Others may be able to infer information about agents when they are selected by a transparent algorithm. In these cases, algorithmic transparency may impose a privacy cost on agents. Future research can investigate algorithmic transparency in the presence of privacy concerns. We believe these are interesting avenues for future research.

{\bibliography{ms}}

\begin{thebibliography}{35}
\providecommand{\natexlab}[1]{#1}
\providecommand{\url}[1]{\texttt{#1}}
\expandafter\ifx\csname urlstyle\endcsname\relax
  \providecommand{\doi}[1]{doi: #1}\else
  \providecommand{\doi}{doi: \begingroup \urlstyle{rm}\Url}\fi

\bibitem[Alon et~al.(2020)Alon, Dobson, Procaccia, Talgam-Cohen, and
  Tucker-Foltz]{Alon2020}
T.~Alon, M.~R.~C. Dobson, A.~D. Procaccia, I.~Talgam-Cohen, and
  J.~Tucker-Foltz.
\newblock Multiagent evaluation mechanisms.
\newblock In \emph{AAAI}, 2020.

\bibitem[Alós-Ferrer and Prat(2012)]{ALOSFERRER2012}
C.~Alós-Ferrer and J.~Prat.
\newblock Job market signaling and employer learning.
\newblock \emph{Journal of Economic Theory}, 147\penalty0 (5):\penalty0 1787 --
  1817, 2012.

\bibitem[Arrieta et~al.(2020)Arrieta, D{\'\i}az-Rodr{\'\i}guez, Del~Ser,
  Bennetot, Tabik, Barbado, Garc{\'\i}a, Gil-L{\'o}pez, Molina, Benjamins,
  et~al.]{arrieta2020explainable}
A.~B. Arrieta, N.~D{\'\i}az-Rodr{\'\i}guez, J.~Del~Ser, A.~Bennetot, S.~Tabik,
  A.~Barbado, S.~Garc{\'\i}a, S.~Gil-L{\'o}pez, D.~Molina, R.~Benjamins, et~al.
\newblock Explainable artificial intelligence (xai): Concepts, taxonomies,
  opportunities and challenges toward responsible ai.
\newblock \emph{Information Fusion}, 58:\penalty0 82--115, 2020.

\bibitem[Bertsimas et~al.(2019)Bertsimas, Delarue, Jaillet, and
  Martin]{bertsimas2019price}
D.~Bertsimas, A.~Delarue, P.~Jaillet, and S.~Martin.
\newblock The price of interpretability.
\newblock \emph{arXiv preprint arXiv:1907.03419}, 2019.

\bibitem[Bonatti and Cisternas(2019)]{Bonatti2019}
A.~Bonatti and G.~Cisternas.
\newblock {Consumer Scores and Price Discrimination}.
\newblock \emph{The Review of Economic Studies}, 87\penalty0 (2):\penalty0
  750--791, 2019.

\bibitem[Brauneis and Goodman(2017)]{brauneis}
R.~Brauneis and E.~P. Goodman.
\newblock Algorithmic transparency for the smart city.
\newblock \emph{20 Yale J. of Law and Tech. 103 (2018); GWU Law School Public
  Law Research Paper; GWU Legal Studies Research Paper}, 2017.

\bibitem[Brückner et~al.(2012)Brückner, Kanzow, and Scheffer]{bruckner2012}
M.~Brückner, C.~Kanzow, and T.~Scheffer.
\newblock Static prediction games for adversarial learning problems.
\newblock \emph{The Journal of Machine Learning Research}, 13:\penalty0
  2617--2654, 09 2012.

\bibitem[Chen et~al.(2018)Chen, Podimata, Procaccia, and Shah]{chen2018}
Y.~Chen, C.~Podimata, A.~D. Procaccia, and N.~Shah.
\newblock Strategyproof linear regression in high dimensions.
\newblock In \emph{Proceedings of the 2018 ACM Conference on Economics and
  Computation}, pages 9--26, 2018.

\bibitem[Citron and Pasquale(2014)]{Citron}
D.~K. Citron and F.~A. Pasquale.
\newblock The scored society: Due process for automated predictions.
\newblock \emph{Washington Law Review}, 89, 2014.

\bibitem[Cowgill and Tucker(2020)]{cowgill2020algorithmic}
B.~Cowgill and C.~E. Tucker.
\newblock Algorithmic fairness and economics.
\newblock \emph{The Journal of Economic Perspectives}, 2020.

\bibitem[Crawford and Sobel(1982)]{crawford1982strategic}
V.~P. Crawford and J.~Sobel.
\newblock Strategic information transmission.
\newblock \emph{Econometrica}, pages 1431--1451, 1982.

\bibitem[Cummings et~al.(2015)Cummings, Ioannidis, and Ligett]{cummings2015}
R.~Cummings, S.~Ioannidis, and K.~Ligett.
\newblock Truthful linear regression.
\newblock In \emph{Conference on Learning Theory}, pages 448--483, 2015.

\bibitem[Daley and Green(2014)]{DALEY2014}
B.~Daley and B.~Green.
\newblock Market signaling with grades.
\newblock \emph{Journal of Economic Theory}, 151:\penalty0 114 -- 145, 2014.

\bibitem[Dalvi et~al.(2004)Dalvi, Domingos, Sanghai, and
  Verma]{adversarialclassification}
N.~Dalvi, P.~Domingos, S.~Sanghai, and D.~Verma.
\newblock Adversarial classification.
\newblock In \emph{Proceedings of the tenth ACM SIGKDD international conference
  on Knowledge discovery and data mining}, pages 99--108, 2004.

\bibitem[Engers(1987)]{engers1987}
M.~Engers.
\newblock Signalling with many signals.
\newblock \emph{Econometrica}, 55\penalty0 (3):\penalty0 663--674, 1987.

\bibitem[Ford and Price(2016)]{Price}
R.~A. Ford and W.~Price.
\newblock Privacy and accountability in black-box medicine.
\newblock \emph{Mich. Telecomm. \& Tech. L. Rev.}, 23:\penalty0 1, 2016.

\bibitem[Frankel and Kartik(2019)]{frankel2019}
A.~Frankel and N.~Kartik.
\newblock Improving information from manipulable data.
\newblock \emph{arXiv preprint arXiv:1908.10330}, 2019.

\bibitem[Fu et~al.(2019)Fu, Huang, and Singh]{fu2018crowd}
R.~Fu, Y.~Huang, and P.~V. Singh.
\newblock Crowd, lending, machine, and bias.
\newblock \emph{Available at SSRN 3206027}, 2019.

\bibitem[Goodhart(1984)]{goodhart1984problems}
C.~A. Goodhart.
\newblock Problems of monetary management: the uk experience.
\newblock In \emph{Monetary Theory and Practice}, pages 91--121. Springer,
  1984.

\bibitem[Goodman and Flaxman(2017)]{goodman2017european}
B.~Goodman and S.~Flaxman.
\newblock European union regulations on algorithmic decision-making and a
  “right to explanation”.
\newblock \emph{AI magazine}, 38\penalty0 (3):\penalty0 50--57, 2017.

\bibitem[Haghtalab et~al.(2020)Haghtalab, Immorlica, Lucier, and
  Wang]{haghtalab2020}
N.~Haghtalab, N.~Immorlica, B.~Lucier, and J.~Wang.
\newblock Maximizing welfare with incentive-aware evaluation mechanisms.
\newblock In \emph{29th International Joint Conference on Artificial
  Intelligence}, 2020.

\bibitem[Hardt et~al.(2016)Hardt, Megiddo, Papadimitriou, and Wootters]{Hardt}
M.~Hardt, N.~Megiddo, C.~Papadimitriou, and M.~Wootters.
\newblock Strategic classification.
\newblock In \emph{Proceedings of the 2016 ACM conference on innovations in
  theoretical computer science}, pages 111--122, 2016.

\bibitem[Kleinberg and Raghavan(2019)]{kleinberg2018}
J.~Kleinberg and M.~Raghavan.
\newblock How do classifiers induce agents to invest effort strategically?
\newblock In \emph{Proceedings of the 2019 ACM Conference on Economics and
  Computation}, pages 825--844, 2019.

\bibitem[Kroll et~al.(2017)Kroll, Huey, Barocas, Felten, Reidenberg, Robinson,
  and Yu]{kroll}
J.~A. Kroll, J.~Huey, S.~Barocas, E.~W. Felten, J.~R. Reidenberg, D.~G.
  Robinson, and H.~Yu.
\newblock Accountable algorithms.
\newblock \emph{University of Pennsylvania Law Review}, 165, 2017.

\bibitem[Lambrecht and Tucker(2019)]{lambrecht2019algorithmic}
A.~Lambrecht and C.~Tucker.
\newblock Algorithmic bias? an empirical study of apparent gender-based
  discrimination in the display of stem career ads.
\newblock \emph{Management Science}, 65\penalty0 (7):\penalty0 2966--2981,
  2019.

\bibitem[Lucas et~al.(1976)]{lucas1976econometric}
R.~E. Lucas et~al.
\newblock Econometric policy evaluation: A critique.
\newblock In \emph{Carnegie-Rochester conference series on public policy},
  volume~1, pages 19--46, 1976.

\bibitem[Meir et~al.(2012)Meir, Procaccia, and Rosenschein]{meir2012}
R.~Meir, A.~Procaccia, and J.~Rosenschein.
\newblock Algorithms for strategyproof classification.
\newblock \emph{Journal of Artificial Intelligence}, 186:\penalty0 123--156, 07
  2012.

\bibitem[Milli et~al.(2019)Milli, Miller, Dragan, and Hardt]{milli2018}
S.~Milli, J.~Miller, A.~D. Dragan, and M.~Hardt.
\newblock The social cost of strategic classification.
\newblock In \emph{Proceedings of the Conference on Fairness, Accountability,
  and Transparency}, pages 230--239, 2019.

\bibitem[Pasquale(2015)]{theblackboxsociety}
F.~A. Pasquale.
\newblock \emph{The Black Box Society: The Secret Algorithms That Control Money
  and Information}.
\newblock Harvard University Press, 2015.

\bibitem[Pearl et~al.(2000)]{pearl2000models}
J.~Pearl et~al.
\newblock \emph{Models, reasoning and inference}.
\newblock Cambridge, UK: Cambridge University Press, 2000.

\bibitem[Schellmann and Bellini(2018)]{schellmann2018artificial}
H.~Schellmann and J.~Bellini.
\newblock Artificial intelligence: The robots are now hiring.
\newblock \emph{The Wall Street Journal}, 20, 2018.

\bibitem[Segal(2011)]{segal2011dirty}
D.~Segal.
\newblock The dirty little secrets of search.
\newblock \emph{The New York Times}, 12\penalty0 (02), 2011.

\bibitem[Spence(1973)]{Spence1973}
M.~Spence.
\newblock Job market signaling.
\newblock \emph{Quarterly Journal of Economics}, 87:\penalty0 355–374, 1973.

\bibitem[Weiss(1983)]{weiss1983sorting}
A.~Weiss.
\newblock A sorting-cum-learning model of education.
\newblock \emph{Journal of Political Economy}, 91\penalty0 (3):\penalty0
  420--442, 1983.

\bibitem[Wladawsky-Berger(2019)]{wladawsky2019current}
I.~Wladawsky-Berger.
\newblock The current state of ai adoption.
\newblock \emph{Wall Street Journal, 08th Oct}, page~1, 2019.

\end{thebibliography}

\newpage
\appendix
\numberwithin{equation}{section}

\section{Mathematical Appendix}

\subsection{Proof of Lemma \ref{lemma1}}\label{Proof of opaque}

In the proof, we will proceed as follows: First, we will study each equilibrium outcome by both identifying the region in the $C_H$-$C_L$ space in which this equilibrium can be realized and analyzing the corresponding payoffs for the firm and agents. 


\vspace{0.1in}\noindent {\bf Opaque Case 1.} We first look at the case where neither \emph{H} type nor \emph{L} agents improve on education. In this case, $\gamma^E_S = \gamma^B_S$. The firm's strategy is to use: $P_{A}=P_{B}=P_{D}=1,P_{C}=0$. To guarantee that this is a Nash equilibrium, we need the following conditions to be hold:
\begin{align*}
& C_{H} \geq (1-\lambda)R       & \text {(\emph{H} type agents will not deviate)} \\
& C_{L} \geq \lambda R          & \text {(\emph{L} type agents will not deviate)} \\
& \gamma^E_{A} \geq \gamma_{th0}  & \text {(the firm will not deviate on $P_{A}$)}\\
& \gamma^E_{C} \leq \gamma_{th0}  & \text {(the firm will not deviate on $P_{C}$)}.
\end{align*}

\noindent
(The first two conditions say that \emph{H} type and \emph{L} type agents are better off (in terms of utility) by not switching to improving on education, and the last two conditions say that the firm is better off (in terms of total payoffs) by not switching its strategies on $P_A$ and $P_C$.)  The first two conditions specify the regions in $C_{H}-C_{L}$ space that can induce this equilibrium (the graph below shows this region). The last two conditions are the direct consequences of Assumption (\ref{as:2}):
\begin{align*}
& \frac{\lambda \theta}{\lambda \theta+(1-\lambda)(1-\theta)} \geq \frac{R}{\alpha} = \gamma_{th0}
& \frac{(1-\lambda) \theta}{(1-\lambda) \theta+\lambda(1-\theta)} \leq \theta \leq \frac{R}{\alpha} = \gamma_{th0}.
\end{align*}

\begin{figure}[htbp]
\centering
\includegraphics[scale=0.4]{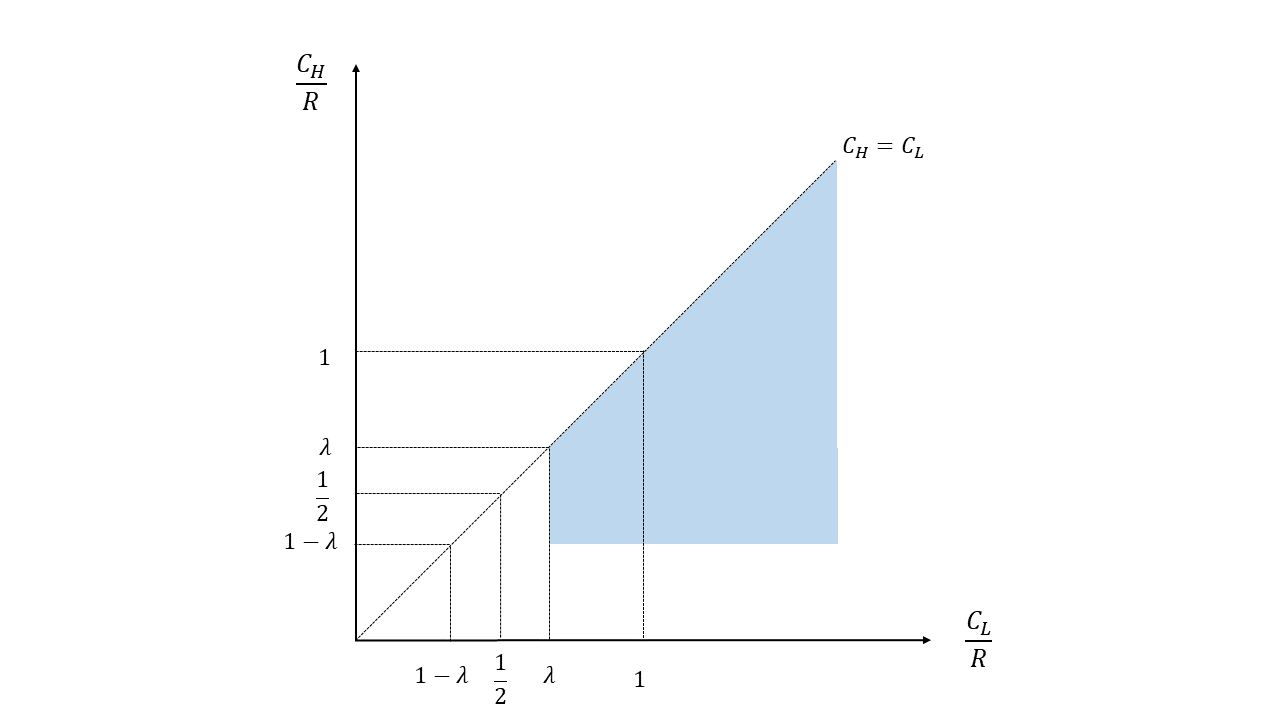}
\caption{opaque case 1}
\label{fig:opaque case 1}
\end{figure}

Total payoffs to each side are given by:
\begin{align*}
& \Pi_{firm_O1}=\lambda \theta \alpha-(\lambda \theta+(1-\lambda)(1-\theta))R \\
& \Pi_{H_O1}=\theta \lambda R \\
& \Pi_{L_O1}=(1-\theta)(1-\lambda)R.
\end{align*}

\noindent
where we use $\Pi_{H_O1}$ and $\Pi_{L_O1}$ to denote the total payoff of \emph{H} type and \emph{L} type agents, respectively. (In the remaining of the proof, we will use $\Pi_{H_Oi}$ and $\Pi_{L_Oi}$ to denote total payoff of \emph{H} type and \emph{L} type agents under case $i$, respectively.)

\vspace{0.1in}\noindent {\bf Opaque Case 2.} In this case, only \emph{H} type agents improve on education and we have: $\gamma^E_{B}=\gamma^E_{D}=1$ and $\gamma^E_{A}=\gamma^E_{C}=0$. We find another equilibrium where the firm's strategy is $P_{A}=P_{C}=0$, $P_{B}=P_{D}=1$. The conditions on the parameters are:

\begin{align*}
& C_{H} \leq R       & \text {(\emph{H} type agents will not deviate)} \\
& C_{L} \geq R          & \text {(\emph{L} type agents will not deviate)} \\
& \gamma^E_{B} \geq \gamma_{th1}, \gamma^E_{D} \geq \gamma_{th1}   & \text {(the firm will not deviate on $P_{B}$ and $P_{D}$)}\\
& \gamma^E_{A} \leq \gamma_{th0}, \gamma^E_{C} \leq \gamma_{th0}   & \text {(the firm will not deviate on $P_{A}$ and $P_{C}$)}.
\end{align*}

The first two conditions specify the regions (see the graph below) that can induce this equilibrium. The last two constraints are trivially satisfied.

\begin{figure}[htbp]
\centering
\includegraphics[scale=0.4]{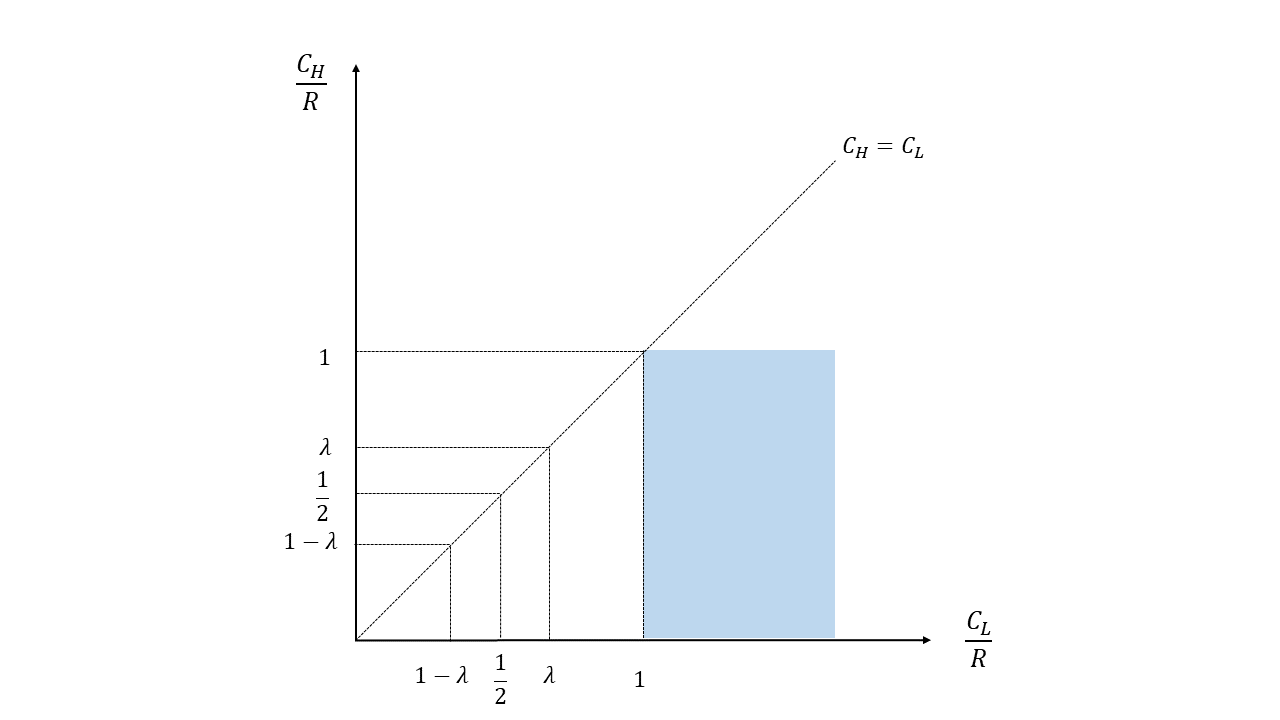}
\caption{opaque case 2}
\label{fig:opaque case 2}
\end{figure}

Total payoffs to each side:
\begin{align*}
& \Pi_{firm_O2}=\theta (\alpha+\beta)-\theta R \\
& \Pi_{H_O2}=\theta (R-C_{H}) \\
& \Pi_{L_O2}=0.
\end{align*}

\vspace{0.1in}\noindent {\bf Opaque Case 3.} In the case, both \emph{H} type and \emph{L} type agents improve on education. The values of $\gamma^E$'s are given below:
\begin{align*}
& \gamma^E_{A}=\gamma^E_{C}=0 \\
& \gamma^E_{B}=\frac{\lambda \theta}{\lambda \theta+(1-\lambda)(1-\theta)} \\
& \gamma^E_{D}=\frac{(1-\lambda) \theta}{(1-\lambda) \theta+\lambda(1-\theta)}.
\end{align*}

The firm's strategy is to use $P_{A}=P_{C}=P_{D}=0, P_{B}=1$. To guarantee that this is a Nash equilibrium, we need the following conditions to hold:
\begin{align*}
& C_{H} \leq \lambda R       & \text {(\emph{H} type agents will not deviate)} \\
& C_{L} \leq (1-\lambda) R          & \text {(\emph{H} type agents will not deviate)} \\
& \gamma^E_{B} \geq \gamma_{th1}  & \text {(the firm will not deviate on $P_{B}$)}\\
& \gamma^E_{D} \leq \gamma_{th1}  & \text {(the firm will not deviate on $P_{D}$)}.
\end{align*}

The first two conditions specify the regions (see graph below) that can induce this equilibrium. The last two conditions are direct consequences of Equation \ref{as:1} and \ref{as:2}. 
\begin{figure}[htbp]
\centering
\includegraphics[scale=0.4]{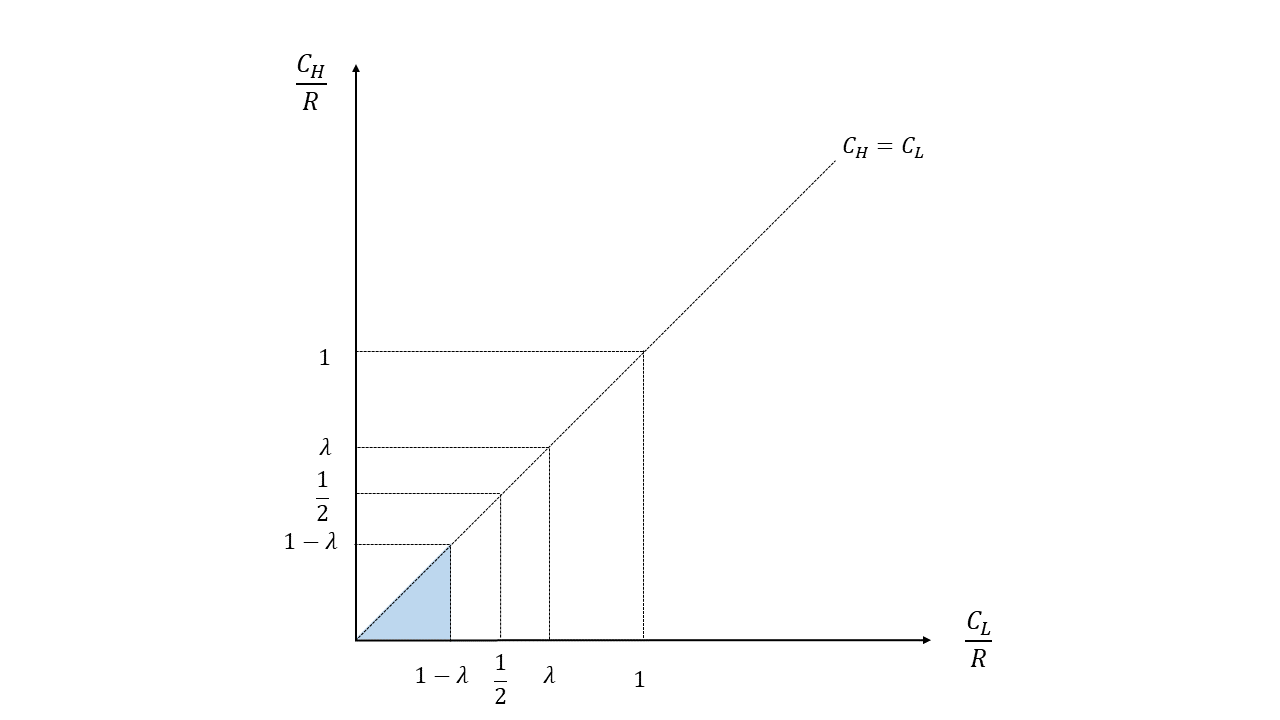}
\caption{opaque case 3}
\label{fig:opaque case 3}
\end{figure}

Total payoffs to each side:
\begin{align*}
& \Pi_{firm_O3}=\lambda \theta (\alpha+\beta)+(1-\lambda)(1-\theta)\beta-(\lambda \theta+(1-\lambda)(1-\theta))R \\
& \Pi_{H_O3}=\theta \lambda R-\theta C_{H} \\
& \Pi_{L_O3}=(1-\theta)(1-\lambda)R-(1-\theta)C_{L}.
\end{align*}


\vspace{0.1in}\noindent {\bf Opaque Case 4.} In this case, \emph{H} type agents improve on education with probability $p_{H}$ and \emph{L} type agents do not improve on education. The values of $\gamma^E$'s are given below:
\begin{align*}
& \gamma^E_{A}=\frac{\theta (1-p_{H})\lambda}{\theta (1-p_{H})\lambda+(1-\theta)(1-\lambda)} \\
& \gamma^E_{C}= \frac{\theta (1-p_{H})(1-\lambda)}{\theta (1-p_{H})(1-\lambda)+(1-\theta)\lambda} \\
& \gamma^E_{B}= \gamma^E_{D}=1.
\end{align*}

The firm's strategy is to use $P_{A}=p_4,P_{C}=0,P_{B}=P_{D}=1$. To guarantee that this is a Nash equilibrium, the following conditions should hold:
\begin{align*}
& C_{H} = (1-\lambda p_4) R       & \text {(\emph{H} type agents are indifferent)} \\
& C_{L} \geq (1-(1-\lambda) p_4) R  & \text {(\emph{L} type agents will not deviate)} \\
& \gamma^E_{A} = \gamma_{th0}     & \text {(the firm is indifferent on $P_A$)}\\
& \gamma^E_{C} \leq \gamma_{th0}  & \text {(the firm will not deviate on $P_C$)}.
\end{align*}

\noindent
The last condition is satisfied following Assumption \ref{as:2}:
\begin{align*}
& \gamma^E_{C}= \frac{\theta(1-p_{H})(1-\lambda)}{\theta(1-p_{H})(1-\lambda)+(1-\theta)\lambda} \leq \frac{(1-\lambda) \theta}{(1-\lambda) \theta+\lambda(1-\theta)} \leq \frac{R}{\alpha} = \gamma_{th0}.
\end{align*}

\noindent
The first condition could be used to represent $p$ in terms of the other parameters and, similarly, the third condition can be used to represent $p_{H}$ in terms of the other parameters. Specifically, we have:
\begin{eqnarray}
p_4 & = & \frac{1}{\lambda} \left(1 - \frac{C_H}{R}\right) \label{eq:p4}\\
p_H & = & 1-\frac{R(1-\theta)(1-\lambda)}{(\alpha-R)\theta \lambda} \label{eq:pH}.
\end{eqnarray}

Given the fact that $p$ and $p_{H}$ are values between 0 and 1, we can calculate the range for $C_{H}$ and $C_{L}$ that lead to this equilibrium (shown in the graph below): 
\begin{align*}
&(1-\lambda)R \leq C_{H} \leq R \\ 
&\frac{R-C_{H}}{R-C_{L}} \leq \frac{\lambda}{1-\lambda}.
\end{align*}

\noindent
where the first inequality follows from the first condition and the second inequality follows from the second condition.

\begin{figure}[htbp]
\centering
\includegraphics[scale=0.4]{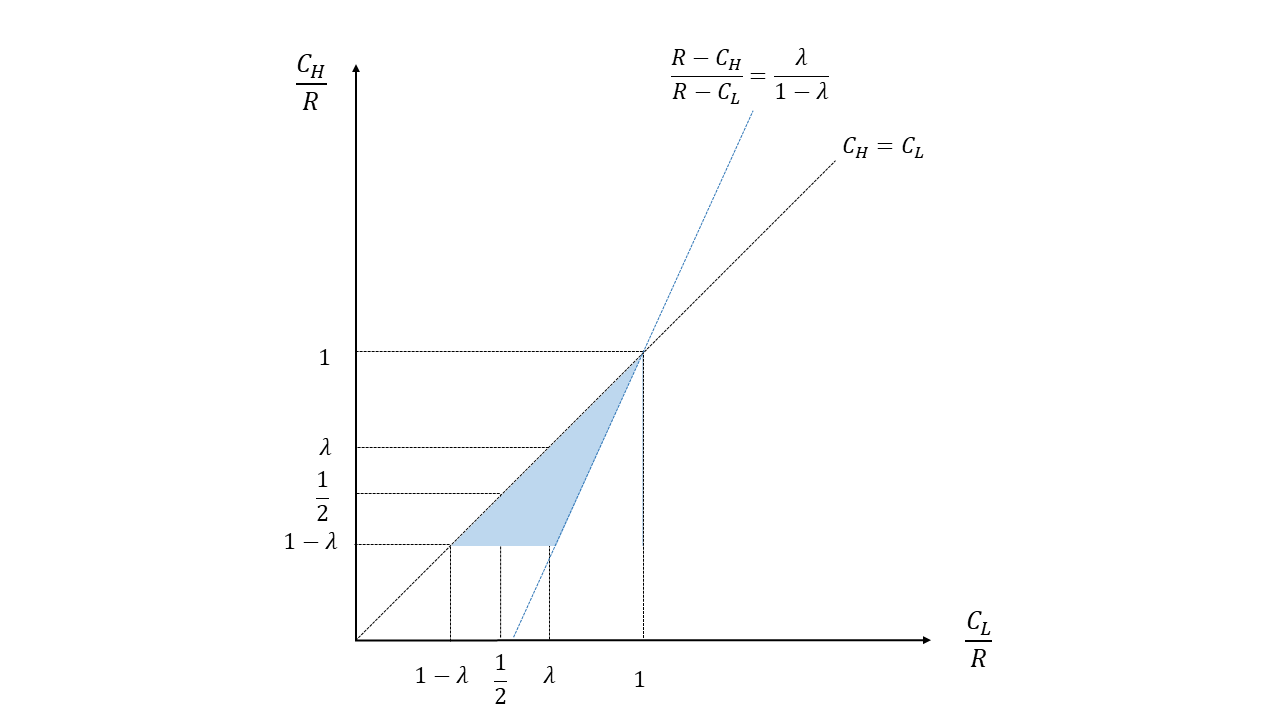}
\caption{opaque case 4}
\label{fig:opaque case 4}
\end{figure}

Total payoffs to each side:
\begin{align*}
& \Pi_{firm_O4}=\theta p_{H}(\alpha+\beta)-\theta p_{H}R\\
& \Pi_{H_O4}=\theta(R-C_{H}) \\
& \Pi_{L_O4}=p_4(1-\theta)(1-\lambda)R.
\end{align*}

\vspace{0.1in}\noindent {\bf Opaque Case 5.} In this case, \emph{H} type agents improve on education and \emph{L} type agents improve on education with probability $p_{L}$. We have:
\begin{align*}
& \gamma^E_{A}= \gamma^E_{C}=0\\
& \gamma^E_{B}=\frac{\theta \lambda}{\theta \lambda+(1-\theta)p_{L}(1-\lambda)} \\
& \gamma^E_{D}= \frac{\theta (1-\lambda)}{\theta (1-\lambda)+(1-\theta) p_{L}\lambda}. 
\end{align*}

The firm's strategy is to use $P_{A}=P_{C}=0,P_{B}=1,P_{D}=p_5$. To guarantee that this is a Nash equilibrium, the following conditions should hold:
\begin{align*}
& C_{H} \leq (\lambda+p_5(1-\lambda)) R   & \text {(\emph{H} type agents will not deviate)} \\
& C_{L} = ((1-\lambda)+ p_5\lambda) R  & \text {(\emph{L} type agents are indifferent)} \\
& \gamma^E_{B} \geq \gamma_{th1}     & \text {(the firm will not deviate on $P_B$)}\\
& \gamma^E_{D} = \gamma_{th1}  & \text {(the firm will not deviate on $P_D$)}.
\end{align*}

The second condition can be used to represent $p$ in terms of the other parameters while the last condition can be used to represent $p_{L}$ in terms of the other parameters. Specifically, 
\begin{eqnarray}
p_5 & = & \frac{C_{L}-R}{\lambda R}+1 \label{eq:p5} \\
p_{L} & = &\frac{\alpha \theta(1-\lambda)-(R-\beta)\theta(1-\lambda)}{(1-\theta)\lambda(R-\beta)} \label{eq:pL}.
\end{eqnarray}

\noindent
The third condition is satisfied following Equation \ref{as:2} and \ref{as:3}:
\begin{align*}
& \gamma^E_{B}=\frac{\theta \lambda}{\theta \lambda+(1-\theta)p_{L}(1-\lambda)} \geq  \gamma_{th0}.
\end{align*}

Given the fact that $p$ and $p_{L}$ are values between 0 and 1, we can calculate the range for $C_{H}$ and $C_{L}$ that lead to this equilibrium: 
\begin{align*}
&(1-\lambda)R \leq C_{L} \leq R \\ 
&\frac{R-C_{H}}{R-C_{L}} \geq \frac{1-\lambda}{\lambda}.
\end{align*}

\begin{figure}[htbp]
\centering
\includegraphics[scale=0.4]{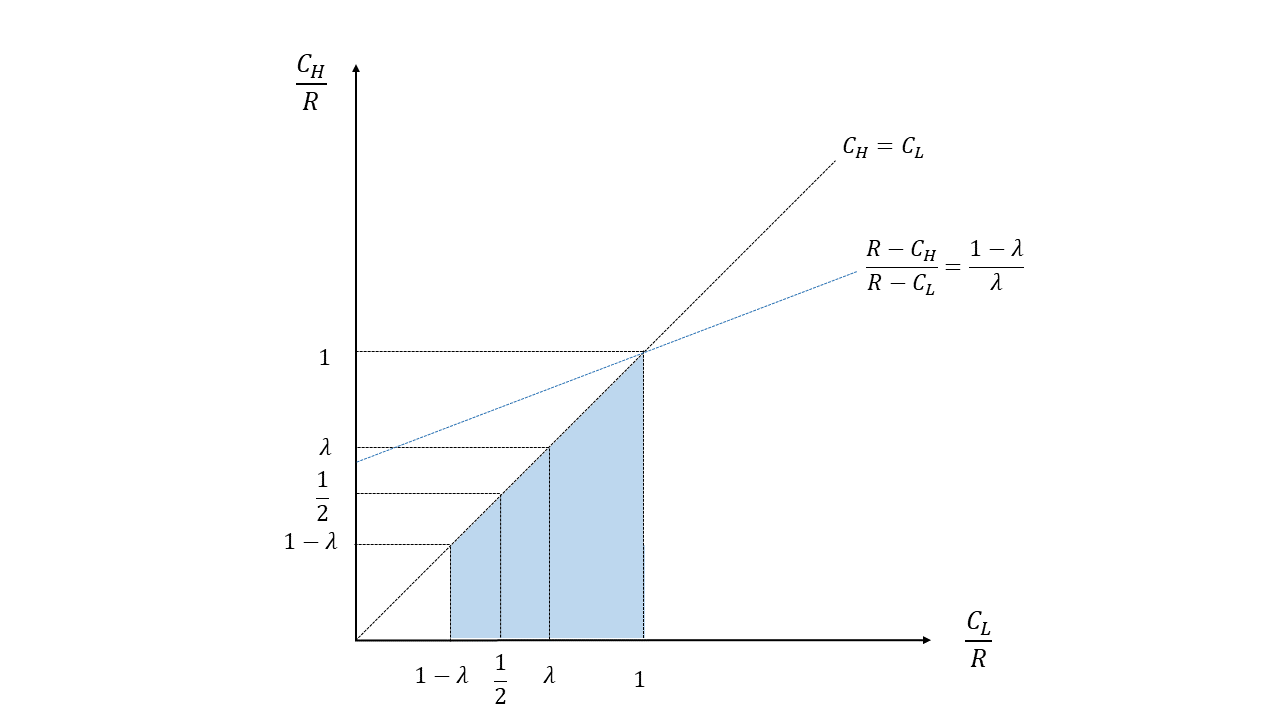}
\caption{opaque case 5}
\label{fig:opaque case 5}
\end{figure}

Total payoffs to each side:
\begin{align*}
\Pi_{firm_O5}&=\theta \lambda(\alpha+\beta-R)+(1-\theta)p_{L}(1-\lambda)(\beta-R)\\
&= \frac{2\lambda-1}{\lambda}\theta(\alpha+\beta-R)\\
\Pi_{H_O5}&=(\theta \lambda+\theta(1-\lambda)p_5)R-\theta C_{H} \\
\Pi_{L_O5}&=0.
\end{align*}

\vspace{0.1in}\noindent {\bf Dealing with multiple equilibria.} Per our analysis of the above five cases, there are several regions where multiple equilibria exist. According to the dynamics of the game, in the opaque case, agents move first and the firm moves next. Thus, the actual equilibrium outcome would be the one gives agents the largest total utilities for each agents' type. (In theory, finding such an equilibrium is not always possible; fortunately, it is in our case.)  
\begin{itemize}
\item In the region where Case 4 and Case 5 overlap, Case 4 always gives higher payoff to both \emph{H} type and \emph{L} type agents:
\begin{align*}
\Pi_{H_O4} & = \theta(R-C_{H}) > (\theta \lambda+\theta(1-\lambda)p_5)R-\theta C_{H} =  \Pi_{H_O5}  \\
\Pi_{L_O4} & = p_{4}(1-\theta)(1-\lambda)R > 0 = \Pi_{L_O5}.
\end{align*}
\noindent
where the inequalities follow since $p_{4}=\frac{R-C_{H}}{\lambda R}$ and $p_{5}=\frac{C_{L}-R}{\lambda R}+1$ are values between $0$ and $1$.

\item In the region where Case 4 and Case 1 overlap, Case 1 always gives higher payoff to both \emph{H} type and \emph{L} type agents:
\begin{align*}
\Pi_{H_O1}& = \theta\lambda R \geq \theta(R-C_{H}) = \Pi_{H_O4}  \\
\Pi_{L_O1}& = (1-\theta)(1-\lambda)R \geq p_{4}(1-\theta)(1-\lambda)R = \Pi_{L_O4}.
\end{align*}
\noindent
where the inequalities follow since  $\frac{C_{H}}{R} \geq 1-\lambda$ in the overlapped region and $p_{4}=\frac{R-C_{H}}{\lambda R}$ is between $0$ and $1$.

\item In the region where Case 1 and Case 2 overlap, Case 1 always gives higher payoff for both \emph{H} type and \emph{L} type agents:
\begin{align*}
\Pi_{H_O1}&= \theta\lambda R \geq \theta(R-C_{H}) =  \Pi_{H_O2}\\
\Pi_{L_O1}&= (1-\theta)(1-\lambda)R \geq 0 =  \Pi_{L_O2}.
\end{align*}
\noindent
where the first inequality follows since  $\frac{C_{H}}{R} \geq 1-\lambda$ in the overlapped region.

\item In the region where Case 1 and Case 5 overlap, Case 1 always gives higher payoff for both \emph{H} type and \emph{L} type agents:
\begin{align*}
\Pi_{H_O1}& \geq \Pi_{H_O4} \geq \Pi_{H_O5} \\
\Pi_{L_O1}& \geq \Pi_{L_O4} \geq \Pi_{L_O5}.
\end{align*}
\noindent
where the first inequality follows since  $\frac{C_{H}}{R} \geq 1-\lambda$ in the overlapped region.
\end{itemize}

\subsection{Proof of Lemma \ref{lemma2}} \label{Proof of transparent}


Similar to the proof of the opaque case, we proceed by analyzing the same five cases analyzed in the opaque case. We show that only the equilibrium outcomes corresponding to cases 1 to 3 are sustainable.   


\vspace{0.1in}\noindent {\bf Transparent Case 1.} In this case, neither \emph{H} type nor \emph{L} type agents improve on education. We have: $\gamma^E_{E}=\theta$ and $\gamma^E_{F}=0$. The firm's strategy is to use $P_{E}=0$ and $P_{F}=1$. To guarantee that this is a Nash equilibrium, the following conditions should hold:
\begin{align*}
& C_{H} \geq  R       & \text {(\emph{H} type agents will not deviate)} \\
& C_{L} \geq  R       & \text {(\emph{L} type agents will not deviate)} \\
& \gamma^E_{E} \leq \gamma_{th0}  & \text {(the firm will not deviate deviate on $P_E$)}.
\end{align*}

The last condition follows from Equation \ref{as:2}:
\begin{align*}
& \gamma^E_{E} = \theta < \frac{R-\beta}{\alpha} < \frac{R}{\alpha} = \gamma_{th0}.
\end{align*}

\begin{figure}[htbp]
\centering
\includegraphics[scale=0.4]{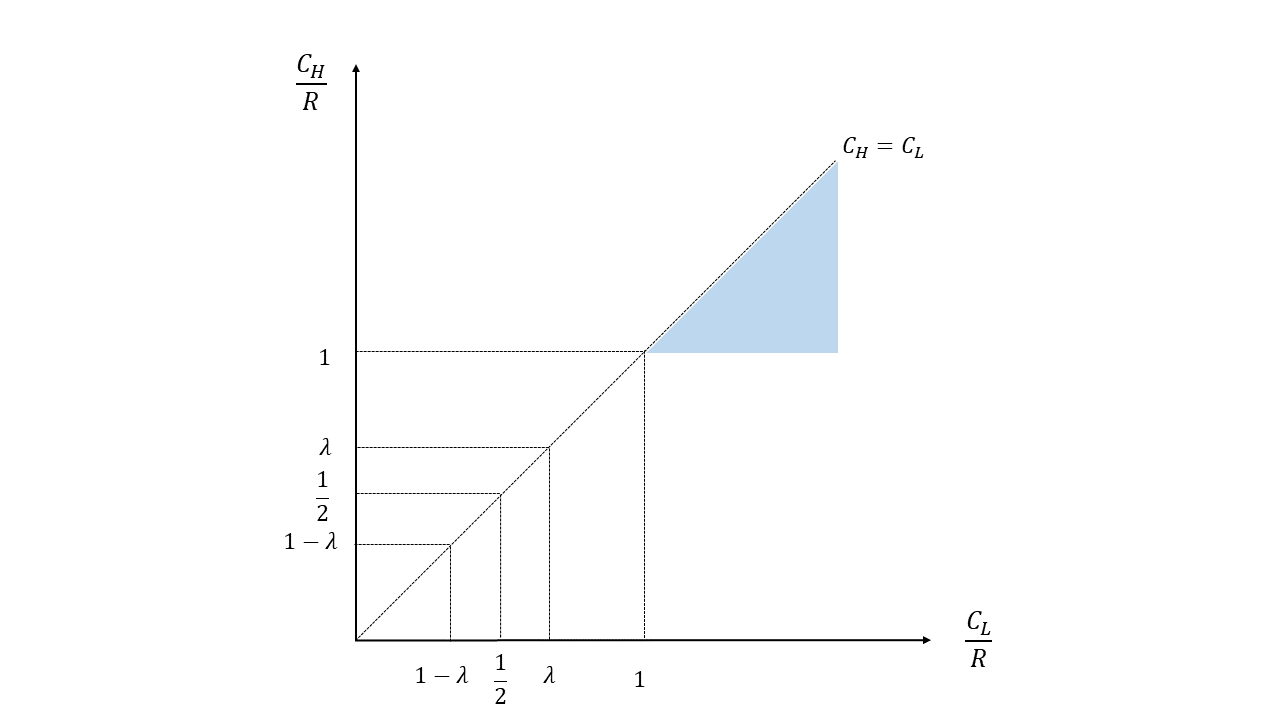}
\caption{transparent case 1}
\label{fig:transparent case 1}
\end{figure}

The region described by the first two conditions on $C_{H}$ and $C_{L}$ is shown in Figure \ref{fig:transparent case 1}. The payoffs to each side:
\begin{align*}
& \Pi_{firm_T1}=0 \\
& \Pi_{H_T1}=0 \\
& \Pi_{L_T1}=0.
\end{align*}
\noindent
where we use $\Pi_{H_Ti}$ and $\Pi_{L_Ti}$ to denote total utilities of \emph{H} type and \emph{L} type agents under case $i$, respectively.

\vspace{0.1in}\noindent {\bf Transparent Case 2.} In this case, only \emph{H} type agents improve on education. We have: $\gamma^E_{E}=0$ and $\gamma^E_{F}=1$. The firm's strategy is to use $P_{E}=0$ and $P_{F}=1$. The conditions needed to guarantee that this is a Nash equilibrium are: 
\begin{align*}
& C_{H} \leq  R       & \text {(\emph{H} type agents will not deviate)} \\
& C_{L} \geq  R       & \text {(\emph{L} type agents will not deviate)} \\
& \gamma^E_{E} \leq \gamma_{th0}     & \text {(the firm will not deviate on $P_E$)}\\
& \gamma^E_{F} \geq \gamma_{th1}  & \text {(the firm will not deviate on $P_F$)}.
\end{align*}

\begin{figure}[htbp]
\centering
\includegraphics[scale=0.4]{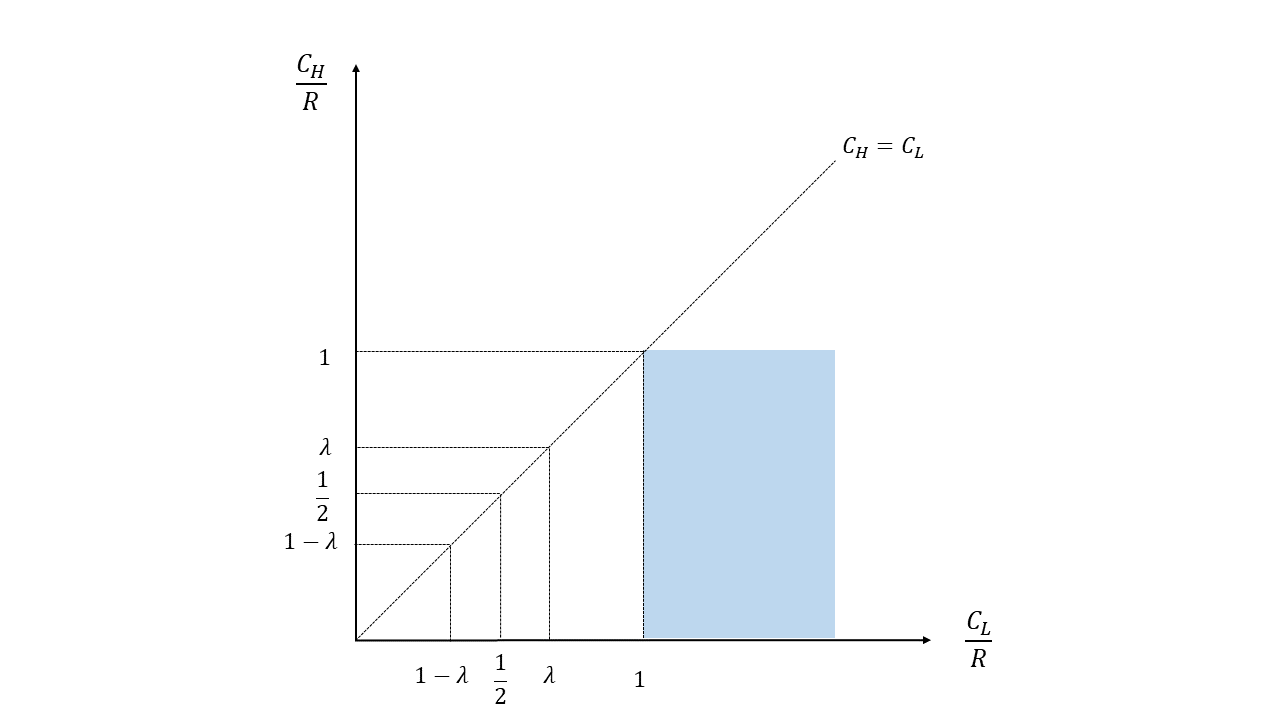}
\caption{transparent case 2}
\label{fig:transparent case 2}
\end{figure}

The last two conditions are trivially satisfied (by Assumption \ref{as:1}, we have $0<\gamma_{th1}<\gamma_{th0}<1$.) The region described by the first two conditions above is shown in Figure \ref{fig:transparent case 2}. The payoffs to each side:
\begin{align*}
& \Pi_{firm_T2}=\theta(\alpha+\beta-R) \\
& \Pi_{H_T2}=\theta(R-C_{H}) \\
& \Pi_{L_T2}=0.
\end{align*}

\vspace{0.1in}\noindent {\bf Transparent Case 3.} In this case, both \emph{H} type and \emph{L} type agents improve on education. We have: $\gamma^E_{E}=0$ and $\gamma^E_{F}=\theta$. The firm's strategy is to use  $P_{E}=0$ and $P_{F}=1$. The conditions needed to guarantee that this is a Nash equilibrium are:
\begin{align*}
& C_{H} \leq  R       & \text {(\emph{H} type agents will not deviate)} \\
& C_{L} \leq  R       & \text {(\emph{L} type agents will not deviate)} \\
& \gamma^E_{F} \geq \gamma_{th1}  & \text {(the firm will not deviate on $P_F$)}.
\end{align*}

\begin{figure}[htbp]
\centering
\includegraphics[scale=0.4]{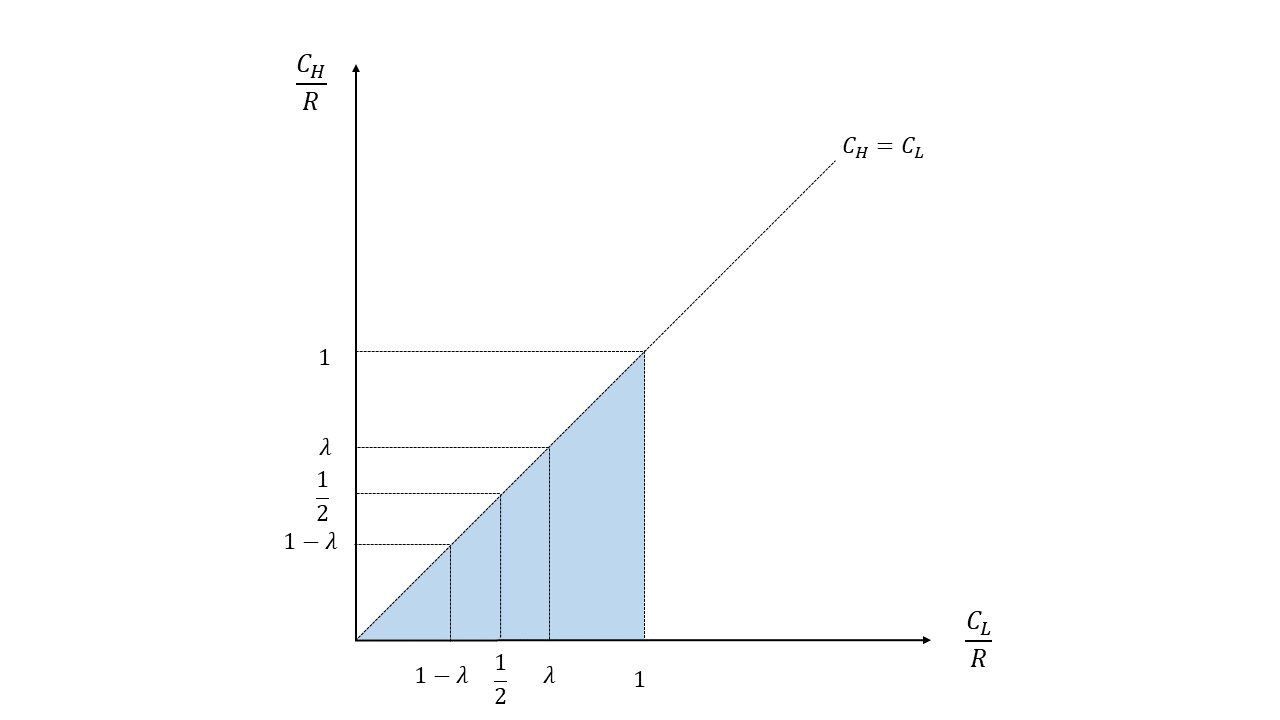}
\caption{transparent case 3}
\label{fig:transparent case 3}
\end{figure}

The third condition follows by Equation \ref{as:3}. The first two conditions specify the regions in the $C_{H}-C_{L}$ space as shown in Figure \ref{fig:transparent case 3}. The payoffs to each side:
\begin{align*}
& \Pi_{firm_T3}=\theta(\alpha+\beta)+(1-\theta)\beta-R \\
& \Pi_{H_T3}=\theta(R-C_{H}) \\
& \Pi_{L_T3}=(1-\theta)(R-C_{L}).
\end{align*}

\vspace{0.1in}\noindent {\bf Transparent Case 4.} In this case, \emph{H} type agents improve on education with probability $p_{H}$ and \emph{L} type agents do not improve on education. We have:
\begin{eqnarray*}
\gamma^E_{E} & = & \frac{(1-p_{H})\theta}{(1-p_{H})\theta+(1-\theta)} \\
\gamma^E_F & = & 1.
\end{eqnarray*}
\noindent
The firm's strategy is to use $P_{E}=p$ and $P_{F}=1$. The conditions needed to guarantee that this is a Nash equilibrium are:
\begin{align*}
& C_{H} =  (1-p)R       & \text {(\emph{H} type agents are indifferent)} \\
& C_{L} \geq  (1-p)R       & \text {(\emph{L} type agents will not deviate)} \\
& \gamma^E_{E} = \gamma_{th0}  & \text {(the firm is indifferent on $P_E$)}.
\end{align*}

Since $p_{H}$ is between 0 and 1, $0<\gamma^E_{E}<\theta$. The last condition requires $0<\frac{R}{\alpha}<\theta$, or equivalently $\alpha>\frac{R}{\theta}$. In the range of $\alpha$ that we are considering (i.e., $\frac{(\theta \lambda+(1-\theta)(1-\lambda))R}{\theta \lambda}<\alpha<\frac{R}{\theta}$, by Assumption \ref{as:2}), this equilibrium cannot be sustained.

\vspace{0.1in}\noindent {\bf Transparent Case 5.} In this case, \emph{H} type agents improve on education and \emph{L} type agents improve on education with probability $p_{L}$. We have:
\begin{eqnarray*}
\gamma^E_{E} & = & 0 \\
\gamma^E_{F} & = & \frac{\theta}{\theta+(1-\theta)p_{L}}.
\end{eqnarray*}

The firm's strategy is to use $P_{E}=0$ and $P_{F}=p$. The conditions needed to guarantee that this is a Nash equilibrium are:
\begin{align*}
& C_{H} \leq  pR       & \text {(\emph{H} type agents will not deviate)} \\
& C_{L} =  pR       & \text {(\emph{L} type agents are indifferent)} \\
& \gamma^E_{F} = \gamma_{th1}  & \text {(the firm is indifferent on $P_F$)}.
\end{align*}

Note that, the second condition implies $p = \frac{C_L}{R}$. Given that $p$ is between 0 and 1, any value of $C_{L}$ between $0$ and $R$ is valid. As for the last condition, $p_{L}$ is between 0 and 1 implies $\theta<\gamma^E_{F}<1$. But, by Assumption \ref{as:3}, $\beta>R-\theta \alpha$, which implies $\gamma_{th1}=\frac{R-\beta}{\alpha}<\theta$. Thus, the last condition cannot be satisfied and, therefore, this equilibrium cannot be sustained. 

\subsection{Derivation of the lower and upper bound of $\alpha$ and $\beta$}\label{Proof of alpha and beta}
In this paper we assume $\alpha$ to be in a certain range to eliminate uninteresting scenarios: 

 $$\frac{(\theta \lambda+(1-\theta)(1-\lambda))R}{\theta \lambda}<\alpha<\frac{R}{\theta}.$$

In the opaque case, when there is no agent improves on the causal feature, we want the firm to hire some agents based on the information in the correlational feature instead of not hiring anyone. (not hiring anyone in this case is uninteresting because it will trivially drive everyone improving on the causal feature). Thus we want $\alpha$ to be large enough to incentivize the firm hiring agents who have value 1 on the correlational feature. In the transparent case, when there is no agent improves on the causal feature, all the agents are mixed together in the feature space: they all have the same values on both the causal and correlational feature. The firm will either hire everyone or not hire anyone, depending one whether the average productivity of all the agents exceeds the salary or not. We want $\alpha$ to be small enough that the firm will not hire anyone in this case. (hiring everyone in this case is uninteresting because no one will have incentive to improve on the causal feature regardless of the cost of improving). Specifically, rewrite the left inequality as $\theta\lambda\alpha+(1-\theta)(1-\lambda)\times 0 > (\theta \lambda+(1-\theta)(1-\lambda))R$. In the initial distribution of the opaque case, (i.e., where everyone has a value of 0 on the causal feature), there are $\theta\lambda$ \emph{H} type agents and $(1-\theta)(1-\lambda)$ \emph{L} type agents who have value 0 on the causal feature and value 1 on the correlational feature. The inequality means that their total productivity (left hand side) should be larger than the total salary paid to them (right hand side). In other words, the firm has an incentive to hire all of these agents. If this is not the case, then the firm will not hire anyone with value 0 on the causal feature even if they have value 1 on the correlational feature which will trivially incentivize individuals to improve on the causal features. The right inequality means in the transparent case where the correlational feature is gamed, if everyone has value 0 on the causal feature, the firm will not hire anyone.

In this paper we also assume $\beta$ to be in a certain range to eliminate uninteresting scenarios: 

 $$R-\theta \alpha<\beta<R-\frac{\theta(1-\lambda)\alpha}{\theta(1-\lambda)+(1-\theta)\lambda}.$$

$\beta$ is the marginal effect of education on the agent's productivity. 

Rewrite the left part in-equation as $\theta(\alpha+\beta)+(1-\theta)\beta > (\theta+(1-\theta))R$. In the transparent case where everyone games on the correlational feature and everyone improves on the causal feature, there are $\theta$ \emph{H} type agents and $1-\theta$ \emph{L} type agents who have value 1 on both feature. The inequality means their total productivity (left hand side) is larger than the total salary paid to them (right hand side). In other words, the firm will have incentive to hire all of them. If this is not the case, then in transparent scenario no one will improve on the causal feature and the firm will end up hiring no one. As for the right part in-equation, rewrite it as $\theta(1-\lambda)(\alpha+\beta)+(1-\theta)\lambda\beta < (\theta(1-\lambda)+(1-\theta)\lambda)R$. In the opaque case where no one games on the correlational feature but everyone improves the causal feature, there are $\theta(1-\lambda)$ \emph{H} type agents and $(1-\theta)\lambda$ \emph{L} type agents who have value 1 on the causal feature but value 0 on the causal feature. This in-equation means their total productivity (left hand side) is smaller than the total wage paid to them (right hand side). in other words, the firm will have no incentive to hire anyone of them. If this is not the case, then in the opaque case improving on the causal feature will ensure an agent to be hired regardless of his value on the correlational feature, which will again, lead to an uninteresting equilibrium.

\end{document}